\documentclass[12pt,american,english,intlimits]{article}
\usepackage[T1]{fontenc}
\usepackage[latin9]{inputenc}
\usepackage[a4paper]{geometry}
\usepackage{fancyhdr}
\usepackage{array}
\usepackage{prettyref}
\usepackage{float}
\usepackage{multirow}
\usepackage{amssymb}
\usepackage{graphicx}
\usepackage{setspace}
\usepackage{esint}
\usepackage[authoryear]{natbib}

\makeatletter

\providecommand{\tabularnewline}{\\}

\newenvironment{lyxcode}
{\par\begin{list}{}{
\setlength{\rightmargin}{\leftmargin}
\setlength{\listparindent}{0pt}
\raggedright
\setlength{\itemsep}{0pt}
\setlength{\parsep}{0pt}
\normalfont\ttfamily}%
 \item[]}
{\end{list}}

\usepackage{psfrag}
\usepackage{units}

\newrefformat{cap}{\hyperref[#1]{Figure~\ref{#1}}}
\newrefformat{fig}{\hyperref[#1]{Figure~\ref{#1}}}
\newrefformat{sec}{\hyperref[#1]{Section~\ref{#1}}}
\newrefformat{sub}{\hyperref[#1]{Section~\ref{#1}}}
\newrefformat{tab}{\hyperref[#1]{Table \ref{#1}}}
\newrefformat{eq}{\hyperref[#1]{Equation~\ref{#1}}}

\@ifundefined{definecolor}
 {\usepackage{color}}{}

\definecolor{parametergray}{gray}{0.8}

\usepackage[OT1]{fontenc}

\renewenvironment{abstract}
{\noindent{\normalfont\large\textbf{Abstract}}%
\par\vspace{0.5\baselineskip}\noindent}
{\par}

\renewcommand{\@seccntformat}[1]{%
\csname the#1\endcsname\hspace{0.5em}}

\renewcommand{\section}{\@startsection
{section}%
{1}%
{0mm}%
{-\baselineskip}%
{0.5\baselineskip}%
{\normalfont\large\bfseries}}

\renewcommand{\subsection}{\@startsection
{subsection}%
{1}%
{0mm}%
{-\baselineskip}%
{0.5\baselineskip}%
{\normalfont\bfseries}}

\renewcommand{\subsubsection}{\@startsection
{subsubsection}%
{2}%
{1em}%
{-\baselineskip}%
{-\fontdimen2\font plus -\fontdimen3\font minus -\fontdimen4\font}%
{\normalfont\bfseries}}

\usepackage{babel}

\usepackage[
colorlinks=true,linkcolor=blue,citecolor=blue,urlcolor=blue,filecolor=blue,
pdffitwindow,
pdftitle={Is a 4-bit synaptic weight resolution enough?},
pdfauthor={Thomas Pfeil},
pdfsubject={},
pdfkeywords={}
]{hyperref}

\usepackage{colortbl}
\usepackage{multirow}
\usepackage{rotate}

\@ifundefined{showcaptionsetup}{}{%
 \PassOptionsToPackage{caption=false}{subfig}}
\usepackage{subfig}
\makeatother

\usepackage{babel}
\begin{document}

\begin{titlepage}\thispagestyle{empty}\setcounter{page}{0}\pdfbookmark[1]{Title}{TitlePage}

\begin{center}
\textbf{\LARGE{Is a 4-bit synaptic weight resolution enough? -- Constraints
on enabling spike-timing dependent plasticity in neuromorphic hardware}}
\par\end{center}{\LARGE \par}

\begin{center}
\textbf{\large{Thomas Pfeil{*}$^{1}$, Tobias C. Potjans$^{2,3}$,
Sven Schrader$^{1}$, Wiebke Potjans$^{2,4}$, Johannes Schemmel$^{1}$,
Markus Diesmann$^{2,3,4,5}$, Karlheinz Meier$^{1}$}}
\par\end{center}{\large \par}

\vfill{}
\singlespacing

\begin{flushleft}
$^{1}$\parbox[t]{12cm}{Kirchhoff Institute for Physics\\
Ruprecht-Karls-University Heidelberg\\
Heidelberg, Germany\raisebox{11cm}[0cm][0cm]{\hspace*{12.0cm}\textbf{\ }}}\\[6mm]
\par\end{flushleft}

\begin{flushleft}
$^{2}$\parbox[t]{12cm}{Institute of Neuroscience and Medicine\\
Computational and Systems Neuroscience (INM-6)\\
Research Center Jülich,\\
Jülich, Germany\raisebox{11cm}[0cm][0cm]{\hspace*{12.0cm}\textbf{\ }}}\\[6mm]
\par\end{flushleft}

\begin{flushleft}
$^{3}$\parbox[t]{12cm}{Brain and Neural Systems Team\\
RIKEN Computational Science Research Program\\
Wako-shi, Japan\raisebox{11cm}[0cm][0cm]{\hspace*{12.0cm}\textbf{\ }}}\\[6mm]
\par\end{flushleft}

\begin{flushleft}
$^{4}$\parbox[t]{12cm}{RIKEN Brain Science Institute\\
Wako-shi, Japan\raisebox{11cm}[0cm][0cm]{\hspace*{12.0cm}\textbf{\ }}}\\[6mm]
\par\end{flushleft}

\begin{flushleft}
$^{5}$\parbox[t]{12cm}{Medical Faculty\\
RWTH Aachen University\\
Aachen, Germany\raisebox{11cm}[0cm][0cm]{\hspace*{12.0cm}\textbf{\ }}}\\[6mm]
\par\end{flushleft}

\vfill{}
\noindent{*} Correspondence:\hspace{1em}\parbox[t]{11cm}{Thomas
Pfeil

Ruprecht-Karls-University Heidelberg\\
Kirchhoff Institute for Physics

Im Neuenheimer Feld 227\\
69120 Heidelberg, Germany\\
tel: +49-6221-549813\\
 \href{mailto:thomas.pfeil@kip.uni-heidelberg.de}{thomas.pfeil@kip.uni-heidelberg.de}}

\end{titlepage}

\begin{abstract}\pdfbookmark[1]{Abstract}{AbstractPage}

Large-scale neuromorphic hardware systems typically bear the trade-off
between detail level and required chip resources. Especially when
implementing spike-timing-dependent plasticity, reduction in resources
leads to limitations as compared to floating point precision. By design,
a natural modification that saves resources would be reducing synaptic
weight resolution. In this study, we give an estimate for the impact
of synaptic weight discretization on different levels, ranging from
random walks of individual weights to computer simulations of spiking
neural networks. The FACETS wafer-scale hardware system offers a 4-bit
resolution of synaptic weights, which is shown to be sufficient within
the scope of our network benchmark. Our findings indicate that increasing
the resolution may not even be useful in light of further restrictions
of customized mixed-signal synapses. In addition, variations due to
production imperfections are investigated and shown to be uncritical
in the context of the presented study. Our results represent a general
framework for setting up and configuring hardware-constrained synapses.
We suggest how weight discretization could be considered for other
backends dedicated to large-scale simulations. Thus, our proposition
of a \textit{good hardware verification practice} may rise synergy
effects between hardware developers and neuroscientists.

\bigskip{}


\end{abstract}

\clearpage{}
\begin{lyxcode}

\end{lyxcode}
\global\long\def\s{\,\mathrm{s}}
\global\long\def\ms{\,\mathrm{ms}}
\global\long\def\kHz{\,\mathrm{kHz}}
\global\long\def\Hz{\,\mathrm{Hz}}
\global\long\def\mV{\,\mathrm{mV}}
\global\long\def\pA{\,\mathrm{pA}}
\global\long\def\nS{\,\mathrm{nS}}
\global\long\def\pF{\,\mathrm{pF}}
\global\long\def\ssps{\,\mathrm{SSPs}}
\global\long\def\bits{\,\mathrm{bits}}
\global\long\def\V{\,\mathrm{V}}
\global\long\def\uA{\,\mathrm{\mathrm{\mu A}}}

\global\long\def\ti{t_{\mathrm{i}}}
\global\long\def\tj{t_{\mathrm{j}}}
\global\long\def\cp{c_{\mathrm{p}}}
\global\long\def\cd{c_{\mathrm{d}}}
\global\long\def\nnc{N_{\mathrm{C}}}
\global\long\def\nnu{N_{\mathrm{U}}}
\global\long\def\nnt{N_{\mathrm{T}}}

\global\long\def\taur{\tau_{\mathrm{ref}}}
\global\long\def\taus{\tau_{\mathrm{syn}}}
\global\long\def\taustdp{\tau_{\mathrm{STDP}}}
\global\long\def\gl{g_{\mathrm{L}}}
\global\long\def\el{E_{\mathrm{L}}}
\global\long\def\ee{E_{\mathrm{e}}}
\global\long\def\vreset{V_{\mathrm{reset}}}
\global\long\def\gmax{g_{\mathrm{max}}}
\global\long\def\cm{C_{\mathrm{m}}}
\global\long\def\taum{\tau_{\mathrm{m}}}

\global\long\def\nuc{\nu_{\mathrm{c}}}
\global\long\def\nuw{\nu_{\mathrm{w}}}
\global\long\def\pp{p_{\mathrm{p}}}
\global\long\def\pd{p_{\mathrm{d}}}
\global\long\def\deltats{\Delta t_{\mathrm{s}}}
\global\long\def\wwd{w_{\mathrm{d}}}
\global\long\def\wwc{w_{\mathrm{c}}}
\global\long\def\deltawd{\Delta w_{\mathrm{d}}}
\global\long\def\deltawp{\Delta w_{\mathrm{p}}}
\global\long\def\assp{a_{\mathrm{SSP}}}
\global\long\def\aa{a_{\mathrm{a}}}
\global\long\def\ac{a_{\mathrm{c}}}
\global\long\def\ath{a_{\mathrm{th}}}
\global\long\def\wi{w_{\mathrm{i}}}
\global\long\def\pdel{p_{\mathrm{del}}}
\global\long\def\deltatdel{\Delta t_{\mathrm{del}}}

\global\long\def\deltaton{\Delta t_{\mathrm{on}}}
\global\long\def\aaabs{A_{\mathrm{abs}}}
\global\long\def\aat{A_{\mathrm{t}}}
\global\long\def\aaa{A_{\mathrm{a}}}
\global\long\def\aac{A_{\mathrm{c}}}
\global\long\def\sigmaa{\sigma_{\mathrm{a}}}
\global\long\def\sigmat{\sigma_{\mathrm{t}}}

\global\long\def\pfn#1#2{\psfrag{#1}[cc][cc]{\textsf{#2}}}

\global\long\def\pfnx#1#2{\psfrag{#1}[tc][tc]{\textsf{#2}}}

\global\long\def\pfny#1#2{\psfrag{#1}[bc][bc]{\textsf{#2}}}

\global\long\def\pft#1#2{\pfn{#1}{\scriptsize{#2}}}

\global\long\def\pftx#1#2{\pfnx{#1}{\scriptsize{#2}}}

\global\long\def\pfty#1#2{\pfny{#1}{\scriptsize{#2}}}

\global\long\def\pfb#1#2{\pfn{#1}{\textbf{\small{#2}}}}

\section{Introduction}

Computer simulations have become an important tool to study cortical
networks \citep[e.g.][]{Brunel00_183,Morrison05a,Vogels05a,Markram06_153,Brette07_349,Johanssson07,Morrison07_1437,Kunkel11_00160,Yger2011_229}.
While they provide insight into activity dynamics that can not otherwise
be measured \textit{in vivo} or calculated analytically, their computation
times can be very time-consuming and consequently unsuitable for statistical
analyses, especially for learning neural networks \citep{Morrison07_1437}.
Even the ongoing enhancement of the von Neumann computer architecture
is not likely to reduce simulation runtime significantly, as both
single- and multi-core scaling face their limits in terms of transistor
size \citep{Thompson06_20}, energy consumption \citep{Esmaeilzadeh2011_365_iscas},
or communication \citep{Perrin11_377}.

Neuromorphic hardware systems are an alternative to von Neumann computers
that alleviates these limitations. Their underlying VLSI microcircuits
are especially designed to solve neuron dynamics and can be highly
accelerated compared to biological time \citep{Indiveri11_00073}.
For most neuron models whose dynamics can be analytically stated,
the evaluation of its equations can be determined either digitally
\citep{Plana07_454} by means of numerical methods or with analog
circuits that solve the neuron equations intrinsically \citep{Millner10_nips}.
The analog approach has the advantage of maximal parallelism, as all
neuron circuits are evolving simultaneously in continuous time. Furthermore,
high acceleration factors compared to biological time (e.g.\ up to
$10^{5}$ reported by \citet{Millner10_nips}), can be achieved by
reducing the size of the analog neuron circuits. Nevertheless, many
neuromorphic hardware systems are developed for operation in real-time
to be applied in sensor applications or medical implants \citep{Fromherz02_276,Levi08_241,Vogelstein08_212}.

\begin{figure}
\begin{centering}
\includegraphics[width=85mm]{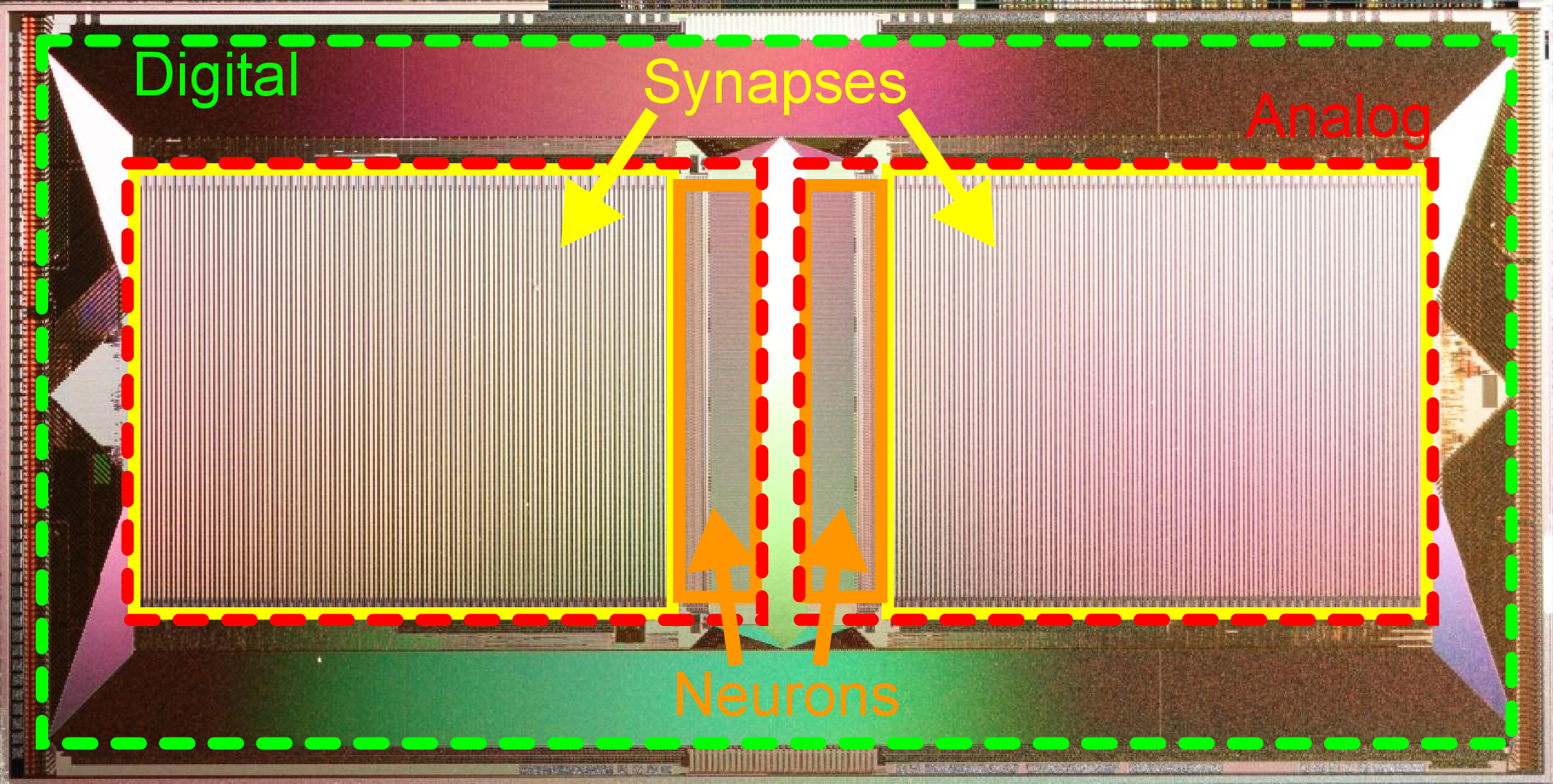}
\par\end{centering}

\caption{Photograph of the HICANN (High Input Count Analog Neural Network)
chip, the basic building block of the FACETS wafer-scale hardware
system. Notice the large area occupied by mixed-signal synapse circuits
(yellow boxes) compared to neuron circuits (orange boxes). A digital
communication infrastructure (area between red and green boxes) ensures
a high density of connections between neurons on the same and to other
HICANN chips.\label{fig:HICANN-photo}}
\end{figure}

Typically, the large number of programmable and possibly plastic
synapses accounts for the major part of chip resources in neuromorphic
hardware systems (\prettyref{fig:HICANN-photo}). Hence, the limited
chip area requires a trade-off between the number and size of neurons
and their synapses, while providing sufficiently complex dynamics.
For example, decreasing the resolution of synaptic weights offers
an opportunity to reduce the area required for synapses and therefore
allows more synapses on a chip, rendering the synaptic weights discretized.

In this study, we will analyze the consequences of such a weight discretization
and propose generic configuration strategies for spike-timing dependent
plasticity on discrete weights. Deviations from original models caused
by this discretization are quantified by particular benchmarks. In
addition, we will investigate further hardware restrictions specific
for the \textit{FACETS}%
\footnote{Fast Analog Computing with Emergent Transient States%
}\textit{ wafer-scale hardware system} \citep{FACETS10}, a pioneering
neuromorphic device that implements a large amount of both configurable
and plastic synapses \citep{Schemmel08_wafer,Schemmel10_1947,Bruederle11_263}.
To this end, custom hardware-inspired synapse models are integrated
into a network benchmark using the simulation tool NEST \citep{Gewaltig_07_11204}.
The objective is to determine the smallest hardware implementation
of synapses without distorting the behavior of theoretical network
models that have been approved by computer simulations.

\clearpage{}

\section{Materials and Methods}

\subsection{Spike-timing dependent plasticity\label{sub:Methods-STDP}}

Here, Spike-Timing Dependent Plasticity (STDP) is treated as a pair-based
update rule as reviewed by e.g.\ \citet{Morrison08_459}. Most pair-based
STDP models \citep{Song00,VanRossum00,Guetig03_3697,Morrison07_1437}
separate weight modifications $\delta w$ into a spike-timing dependent
factor $x(\Delta t)$ and a weight-dependent factor $F(w)$:

\begin{equation}
\delta w(w,\Delta t)=F(w)x(\Delta t),\label{eq:weight_time_separation}
\end{equation}
where $\Delta t=\ti-\tj$ denotes the interval between spike times
$\tj$ and $\ti$ at the pre- and postsynaptic terminal, respectively.
Typically, $x(\Delta t)$ is chosen to be exponentially decaying \citep[e.g.][]{Gerstner96,Kempter99}.

In contrast, the weight-dependence $F(w)$, which is divided into
$F_{+}(w)$ for a causal and $F_{-}(w)$ for an anti-causal spike-timing-dependence,
differs between different STDP models. Examples are given in \prettyref{tab:stdp_models}.
As $F_{+}(w)$ is positive and $F_{-}(w)$ negative for all these
STDP models, causal relationships ($\Delta t>0$) between pre- and
postsynaptic spikes potentiate and anti-causal relationships ($\Delta t<0$)
depress synaptic weights.

In this study, the \textit{intermediate Gütig STDP model} (bounded
to the weight range {[}0,1{]}) is chosen as an example STDP model.
It represents a mixture of the multiplicative ($\mu=1$) and additive
($\mu=0$) STDP model and has been shown to provide stability in competitive
synaptic learning \citep{Guetig03_3697}. Nevertheless, the following
studies can be applied to any pair-based STDP model with exponentially
decaying time-dependence, e.g.\ all models listed in \prettyref{tab:stdp_models}.

\begin{table}
\begin{centering}
\begin{tabular}{>{\raggedright}p{4.5cm}|>{\centering}p{1.9cm}|>{\centering}m{1.9cm}|>{\raggedright}m{1.9cm}}
Model name & $F_{+}(w)$ & $F_{-}(w)$ & $x(\Delta t)$\tabularnewline
\hline 
\hline 
\textbf{Additive}

\citep{Song00} & $\lambda$ & $-\lambda\alpha$ & \tabularnewline
\cline{1-3} 
\textbf{Multiplicative}

\citep{Turrigiano98} & $\lambda(1-w)$ & $-\lambda\alpha w$ & \tabularnewline
\cline{1-3} 
\textbf{Gütig}

\citep{Guetig03_3697} & $\lambda(1-w)^{\mu}$ & $-\lambda\alpha w^{\mu}$ & $\exp(-\frac{|\Delta t|}{\taustdp})$\tabularnewline
\cline{1-3} 
\textbf{Van Rossum}

\citep{VanRossum00} & $\cp$ & $-\cd w$ & \tabularnewline
\cline{1-3} 
\textbf{Power law}

\citep{Morrison07_1437} & $\lambda w^{\mu}$ & $-\lambda\alpha w$ & \tabularnewline
\end{tabular}
\par\end{centering}

\caption{Weight- and spike-timing-dependence of pair-based STDP models: additive,
multiplicative, Gütig, van Rossum and power law model. $F_{+}$ in
case of a causal spike-timing-dependence ($\Delta t>0$) and $F_{-}$
in the anti-causal case ($\Delta t<0$). Throughout this study, the
model proposed by Gütig et al.\  is applied with parameters $\alpha=1.05$,
$\lambda=0.005$, $\mu=0.4$ and $\taustdp=20\ms$ in accordance with
\citet{Song00,VanRossum00,Rubin01,Guetig03_3697,Morrison08_459}.\label{tab:stdp_models}}
\end{table}

\subsection{Synapses in large-scale hardware systems\label{sub:Methods-The-FACETS-Hardware}}

\begin{figure}
\begin{centering}
\includegraphics[width=85mm]{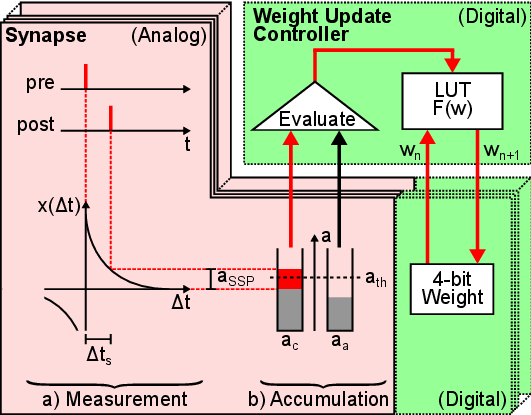}
\par\end{centering}

\caption{Schematic drawing of local hardware synapses which are consecutively
processed by a global weight update controller. Analog circuits are
highlighted in red (with solid frame) and digital circuits in green
(dashed frames). The spike-timing-dependence (here one standard spike
pair (SSP) with $\deltats$, see text) between the pre- and postsynaptic
neuron is \textbf{a)} measured (here $\assp$) and \textbf{b)} accumulated
(here to $\ac$ in case of a causal spike pair, $\aa$ for anti-causal
spike pairs is not affected). Then, the global weight update controller
evaluates the accumulated spike-timing-dependence by means of a crossed
threshold $\ath$ (here $\ac>\ath$) and modifies the digital weight
of the hardware synapse accordingly. The new synaptic weight $w_{\mathrm{n}+1}$
is retrieved from the LUT according to the accumulated spike-timing-dependence
and the current weight $w_{\mathrm{n}}$ and is written back to the
hardware synapse. If either the causal or anti-causal accumulated
spike-timing-dependence crosses the threshold, both accumulations
are reset to zero. The analog measurement and accumulation circuit
is furthermore minimized by using the reduced symmetric nearest-neighbor
spike pairing scheme \citep{Morrison08_459}: instead of considering
all past and future spikes (all-to-all spike pairing scheme), only
the latest and the following spike at both terminals of the synapse
are taken into account.\label{fig:synapse_schematic}}
\end{figure}

The FACETS wafer-scale hardware system \citep{Schemmel08_wafer,Schemmel10_1947,Bruederle11_263}
represents an example for a possible synapse size reduction in neuromorphic
hardware systems. \prettyref{fig:synapse_schematic} schematizes the
hardware implementation of a synapse enabling STDP similar as presented
in \citet{Schemmel06_1} and \citet{Schemmel07_iscas}. It provides
the functionality to store the value of the synaptic weight, to measure
the spike-timing-dependence between pre- and postsynaptic spikes and
to update the synaptic weight according to this measurement. Synapse
density is maximized by separating the \textit{accumulation }of the
spike-timing-dependence $x(\Delta t)$ and the \textit{weight update
controller}, which is the hardware implementation of $F(w)$. This
allows $4\cdot10^{7}$ synapses on a single wafer \citep{Schemmel10_1947}.

Synaptic dynamics in the FACETS wafer-scale hardware system exploits
the fact that weight dynamics typically evolves slower than electrical
neuronal activity \citep{Morrison07_1437,Kunkel11_00160}. Therefore,
weight updates can be divided into two steps (\prettyref{fig:synapse_schematic}).
First, a measuring and accumulation step which locally determines
the relative spike times between pairs of neurons and thus $x(\Delta t)$.
This stage is designed in analog hardware (red area in \prettyref{fig:synapse_schematic}),
as analog measurement and accumulation circuits require less chip
resources compared to digital realizations thereof. Second, the digital
weight update controller (upper green area in \prettyref{fig:synapse_schematic})
implements $F(w)$ based on the previous analog result. A global weight
update controller%
\footnote{One weight update controller for all 256 neurons with 224 synapses
each.%
} is responsible for the consecutive updates of many synapses \citep{Schemmel06_1}
and hence limits the maximal rate at which a synapse can be updated,
the update controller frequency $\nuc$.

Sharing one weight update controller reduces synapses to small analog
measurement and accumulation circuits as well as a digital circuit
that implements the synaptic weight (\prettyref{fig:synapse_schematic}).
The area required to implement these digital weights with a resolution
of $r$ bits is proportional to $2^{r}$, the number of discrete weights.
Consequently, assuming the analog circuits to be fixed in size, the
size of a synapse is determined by its weight storage exponentially
growing with the weight resolution. E.g.\ the FACETS wafer-scale
hardware system has a weight resolution of $r=4\bits$, letting the
previously described circuits (analog and digital) equally sized on
the chip.

Modifications in the layout of synapse circuits are time-consuming
and involve expensive re-manufacturing of chips. Thus, the configuration
of connections between neurons is designed flexible enough to avoid
these modifications and provide a general-purpose modeling environment
\citep{Schemmel10_1947}.\textbf{ }For the same reason, STDP is conform
to the majority of available update rules. The STDP models listed
in \prettyref{tab:stdp_models} share the same time-dependence $x(\Delta t)$.
Its exponential shape is mimicked by small analog circuit not allowing
for other time-dependencies \citep{Schemmel06_1,Schemmel07_iscas}.
The widely differing weight-dependences $F(w)$, on the other hand,
are programmable into the weight update controller. Due to limited
weight update controller resources, arithmetic operations $F(w)$
as listed in \prettyref{tab:stdp_models} are not realizable and are
replaced by a programmable look-up table (LUT) \citep{Schemmel06_1}.

Such a LUT lists, for each discrete weight, the resulting weights
in case of causal or anti-causal spike-timing-dependence between pre-
and postsynaptic spikes. Instead of performing arithmetic operations
during each weight update (\prettyref{eq:weight_time_separation}),
LUTs are used as a recallable memory consisting of precalculated weight
modifications. Hence, LUTs do not limit the flexibility of weight
updates if their weight-dependence (\prettyref{tab:stdp_models})
does not change over time. Throughout this study, we prefer the concept
of LUTs to arithmetic operations, because we like to focus on the
discretized weight space, a state space of limited dimension.

In addition to STDP, the FACETS wafer-scale hardware system also supports
a variant of short-term plasticity mechanisms according to \citet{Tsodyks97}
\citep{Schemmel07_iscas,Bill10_129}, which however leaves synaptic
weights unchanged and therefore lies outside the scope of this study.

\subsection{Discretization of synaptic weights\label{sub:Methods-Discretized-Weights}}

Continuous weight values $\wwc\in[0,1]$, as assumed for the STDP
models listed in \prettyref{tab:stdp_models}, are transformed into
$r$-bit coded discrete weight values $\wwd$:

\begin{equation}
\wwd=c\left\lfloor \frac{\wwc}{c}+\frac{1}{2}\right\rfloor \quad\mathrm{for}\;\wwc\in I\label{eq:conti-dis}
\end{equation}
where $c=1/(2^{r}-1)$ denotes the width of a bin and $\left\lfloor x\right\rfloor $
the floor-function, the largest integer less than or equal to $x$.
This procedure divides the range of weight values $I=[0,1]$ into
$2^{r}$ bins. The term $\frac{1}{2}$ allows for a correct discretization
of weight values near the borders of $I$, effectively dividing the
width of the ending bins (otherwise, only $\wwc=1$ would be mapped
to $\wwd=1$).

\subsection{Discretization of spike-timing dependent plasticity\label{sub:Methods-Look-Up-Tables}}

A single weight update, resulting from a pre- and postsynaptic spike,
might be too fine grained to be captured by a low weight resolution
(\prettyref{eq:conti-dis}). Therefore, it is necessary to accumulate
the effect of weight updates of several consecutive spike pairs in
order to reach the next discrete weight value (\prettyref{eq:conti-dis}
and \prettyref{fig:synapse_schematic}). This is equivalent to state
that the implementation of the STDP model assumes additive features
for ms range intervals. To this end, we define a \textit{standard
spike pair} (SSP) as a spike pair with a time interval between a pre-
and postsynaptic spike of $\deltats=10\ms$ \citep[in accordance to biological measurements by][]{Markram97a,Bi98,Sjostrom01}
in order to provide a standardized measure for the spike-timing-dependence.
This time interval is chosen arbitrarily defining the granularity
only (fine enough for the weight resolutions of interest) and is valid
for both pre-post and post-pre spike pairs, as $x(\Delta t)$ takes
its absolute value.

The values for a LUT are constructed as follows. First, the parameters
$r$ (weight resolution) and $n$ (number of SSPs consecutively applied
for an accumulated weight update) as well as the STDP rule-specific
parameters $\tau_{\mathrm{STDP}}$, $\lambda$, $\mu$, $\alpha$
(\prettyref{tab:stdp_models}) are chosen. Next, starting with a discrete
weight $\wwd$, weight updates $\delta w(w,\deltats)$ specified by
\prettyref{eq:weight_time_separation} are recursively applied $n$
times in continuous weight space using either exclusively $F_{+}(w)$
or $F_{-}(w)$. This results in two accumulated weight updates $\Delta w_{+/-}$,
one for each weight-dependence $F_{+/-}(w)$. Finally, the resulting
weight value in continuous space is according to \prettyref{eq:conti-dis}
transformed back to its discrete representation. This process is then
carried out for each possible discrete weight value $\wwd$ (\prettyref{tab:Example-LUTs}).
We will further compare different LUTs letting $n$ be a free parameter.
In the following a \textit{weight update} refers to $\Delta w$, if
not specified otherwise.

Although we are focusing on the Gütig STDP model, the updated weight
values can in general under- or over-run the allowed weight interval
$I$ due to finite weight updates $\Delta w$. In this case, the weight
is clipped to its minimum or maximum value, respectively.

\begin{table}
\begin{centering}
\renewcommand{\arraystretch}{1.25}%
\begin{tabular}{c|c|c}
$\wwd$ & $w_{+}$ & $w_{-}$\tabularnewline
\hline 
0 & $\frac{1}{3}$ & 0\tabularnewline
$\frac{1}{3}$ & $\frac{2}{3}$ & 0\tabularnewline
$\frac{2}{3}$ & 1 & $\frac{1}{3}$\tabularnewline
1 & 1 & $\frac{2}{3}$\tabularnewline
\end{tabular}
\par\end{centering}

\caption{Example look-up table for a weight resolution of $r=2\bits$ and $n=100$
SSPs. Discrete weight $\wwd$ and the resulting weight increments
$w_{+/-}=\wwd+\Delta w_{+/-}$ for causal and anti-causal weight dependences.\label{tab:Example-LUTs}}
\end{table}

\subsection{Equilibrium weight distributions\label{sub:Methods-Equilibrium-Distributions}}

We analyze long-term effects of weight discretization by studying
the equilibrium weight distribution of a synapse that is subject to
Poissonian pre- and postsynaptic firing. Thus, potentiation and depression
are equally probable ($\pd=\pp=\frac{1}{2}$). Equilibrium weight
distributions in discrete weight space of low resolution (between
2 and 10 bits) are compared to those with high resolution (16 bits)
via the mean squared error $MSE_{eq}$. Consecutive weight updates
are performed based on precalculated LUTs.

Equilibrium weight distributions of discrete weights for a given weight
resolution of $r$ bits are calculated as follows. First, a LUT for
$2^{r}$ discrete weights is configured with $n$ SSPs. Initially,
all $2^{r}$ discrete weight values $\wi$ have the same probability
$P_{\mathrm{i,0}}=\frac{1}{2^{r}}$. For a compact description, the
discrete weights $\wi$ are mapped to a $2^{r}$ dimensional space
with unit vectors $\vec{e}_{\mathrm{i}}\in\mathbb{N}^{2^{r}}$. Then,
for each iteration cycle $j$, the probability distribution is defined
by $\vec{P}_{\mathrm{j}}=\sum_{\mathrm{i=0}}^{2^{r}-1}P_{\mathrm{i,j-1}}(\pp\vec{e}_{\mathrm{c}}+\pd\vec{e}_{\mathrm{a}})$,
where $P_{\mathrm{i,j-1}}$ is the probability for each discrete weight
value $\wi$ of the previous iteration cycle $j-1$. The indices of
$\vec{e}_{\mathrm{c}}$ and $\vec{e}_{\mathrm{a}}$ are those of the
resulting discrete weight values $\wi$ in case of a causal and anti-causal
weight update, respectively, and are represented by the LUT. We define
an equilibrium state as reached if the Euclidean norm $\left\Vert \vec{P}_{\mathrm{j-1}}-\vec{P_{\mathrm{j}}}\right\Vert $
is smaller than a threshold $h=10^{-12}$.

An analytical approach for obtaining equilibrium weight distributions
is derived in \prettyref{sub:Appendix-Analytical-Distributions}.

\subsection{Spiking network benchmarks}

In addition to the behavior under Poissonian noise, we study the impact
of discretized weights with a software implementation of hardware
synapses, enabling us to analyze synapses in isolation as well as
in network benchmarks. The design of our simulation environment is
flexible enough to take further hardware constraints and biological
applications into account.

\subsubsection{Software implementation of hardware synapses\label{sub:Methods-Implementation-NEST}}

The hardware constraints considered in this study are implemented
as a customized synapse model within the framework of the NEST simulation
tool \citep{Gewaltig_07_11204}, allowing their well controlled application
in simulator-based studies on large-scale neural networks. The basic
properties of such a \textit{hardware-inspired synapse} \textit{model}
are described as follows and are illustrated in \prettyref{fig:synapse_schematic}
and \prettyref{fig:sums_and_traces}.

For each LUT configuration defined by its weight resolution $r$ and
number $n$ of SSPs, the threshold for allowing weight updates is
set to 
\begin{equation}
\ath=n\cdot\assp,\label{eq:threshold}
\end{equation}
defining $a=\sum_{\mathrm{i}}x(\Delta t_{\mathrm{i}})$ as the \textit{spike
pair accumulation }for arbitrary intervals. Here, a single SSP is
used, setting $a=\assp=x(\deltats)$. If either the causal or anti-causal
spike pair accumulation $a_{\mathrm{c/a}}$ crosses the threshold
$\ath$, the synapse is ``tagged'' for a weight update. At the next
cycle of the weight update controller all tagged synapses are updated
according to the LUT. Afterwards, the spike pair accumulation (causal
or anti-causal) is reset to zero. Untagged synapses remain unprocessed
by the update controller, and spike pairs are further accumulated
without performing any weight update. If a synapse accumulates $\ac$
and $\aa$ above threshold between two cycles of the weight update
controller, both are reset to zero without updating the synaptic weight.

This threshold process implies that the frequency $\nuw$ of weight
updates is dependent on $n$, which in turn determines the threshold
$\ath$, but also on the firing rates and the correlation between
the pre- and postsynaptic spike train. In general, $a$ increases
faster with higher firing rates or higher correlations. To circumvent
these dependencies on network dynamics, we will use $n$ as a generalized
description for the weight update frequency $\nuw$. The weight update
frequency $\nuw$ should not be confused with the update controller
frequency $\nuc$, with which is checked for threshold crossings and
hence limits $\nuw$.

Furthermore, we have implemented a \textit{reference synapse model}
in NEST, which is based on \citet{Guetig03_3697}. It has the reduction
of employing nearest-neighbor instead of all-to-all spike pairing
\citep{Morrison08_459}.

All simulations involving synapses are simulated with NEST. Spike
trains are applied to built-in \textit{parrot neurons}, that simply
repeat their input, in order to control pre- and postsynaptic spike
trains to interconnecting synapses.

\subsubsection{Single synapse benchmark\label{sub:Methods-Single-Synapse}}

We compare the weight evolutions of hardware-inspired and reference
synapses receiving correlated pre- and postsynaptic spike trains,
drawn from a multiple interaction process (MIP) \citep{Kuhn03_67}.
This process introduces excess synchrony between two realizations
by randomly thinning a template Poisson process. SSPs are then obtained
by shifting one of the processes by $\deltats$. 

In this first scenario the spike pair accumulation $a$ is checked
for crossing $\ath$ with a frequency of $\nuc=10\kHz$ to focus on
the effects of discrete weights only. This frequency is equal to the
simulation step size, preventing the spike pair accumulation from
overshooting the threshold $\ath$ without eliciting a weight update.

Synaptic weights are recorded in time steps of $3\s$ for an overall
period of $150\s$ and are averaged over 30 random MIP realizations.
Afterwards the mean weight at each recorded time step is compared
between the hardware-inspired and the reference synapse model by applying
the mean squared error $MSE_{w}$.

\subsubsection{Network benchmarks\label{sub:Methods-Network-Example}}

The detection of presynaptic synchrony is taken as a benchmark for
synapse implementations. Two populations of $10$ neurons each converge
to an integrate-and-fire neuron with exponentially decaying synaptic
conductances (see schematic in \prettyref{fig:synchrony_detector}A
and model description in \prettyref{tab:Appendix-Model-description}
and \ref{tab:Appendix-Simulation-parameters}) by either hardware-inspired
or reference synapses. These synapses are excitatory, and their initial
weights are drawn randomly from a uniform distribution over $[0,1)$.
The amplitude of the postsynaptic conductance is $w\gmax$ with $\gmax=100\nS$.
One population draws its spikes from a MIP with correlation coefficient
$c$ \citep{Kuhn03_67}, the other from a Poisson process (MIP with
$c\rightarrow0$). We choose presynaptic firing rates of $7.2\Hz$
such that the target neuron settles at a firing rate of $2-22\Hz$
depending on the synapse model . The exact postsynaptic firing rate
is of minor importance as long as the synaptic weights reach an equilibrium
state. The synaptic weights are recorded for $2,000\s$ with a sampling
frequency of $0.1\Hz$. The two resulting weight distributions are
compared applying the Mann-Whitney U test \citet{Mann47_50}.

\paragraph{Further constraints\label{sub:Methods-Further-synapse-constraints}}

Not only the discretization of synaptic weights, but also the update
controller frequency $\nuc$ and the reset behavior are constraints
of the FACETS wafer-scale hardware system.

To study effects caused by a limited update controller frequency,
we choose $\nuc$ such that the interval between sequent cycles is
a multiple of the simulator time step. Consequently weight updates
can only occur on a time grid.

A \textit{common reset} means that both the causal and anti-causal
spike pair accumulations are reset, although only either $\ac$ or
$\aa$ has crossed $\ath$. Because the common reset requires only
one reset line instead of two, it decreases the chip resources of
synapses and is implemented in the current FACETS wafer-scale hardware
system.

As a basis for a possible compensation mechanism for the common reset,
we suggest analog-to-digital converters (ADCs) with a 4-bit resolution
that read out the spike pair accumulations. Such ADCs require only
a small chip area in the global weight update controller compared
to the large area occupied by additional reset lines covering all
synapses and are therefore resource saving alternatives to second
reset lines. An ADC allows to compare the spike pair accumulations
against multiple thresholds. Implementations of the common reset as
well as ADCs are added to the existing software model. For multiple
thresholds, the same number of LUTs is needed that have to be chosen
carefully. To provide symmetry within the order of consecutive causal
and anti-causal weight updates, the spike pair accumulation (causal
or anti-causal) that dominates in means of crossing a higher threshold
is applied first.

\paragraph{Peri-stimulus-time-histograms\label{sub:Methods-Peri-Stimulus-Time-Histograms}}

The difference between static and STDP synapses on eliciting postsynaptic
spikes in the above network benchmark can be analyzed with peri-stimulus-time-histograms
(PSTHs). Here, PSTHs show the probability of postsynaptic spike occurrences
in dependence on the delay between a presynaptic trigger and its following
postsynaptic spike. Spike times are recorded within the last third
of an elongated simulation of $3,000\s$ with $c=0.025$. During the
last $1,000\s$ the mean weights are already in their equilibrium
state, but are still fluctuating around it. The first spike of any
two presynaptic spikes within a time window of $\deltaton=1\ms$ is
used as a trigger. The length of $\deltaton$ is chosen small compared
to the membrane time constant $\taum=15\ms$, such that the excitatory
postsynaptic potentials of both presynaptic spikes overlap each other
and increase the probability of eliciting a postsynaptic spike. On
the other hand $\deltaton$ is chosen large enough to not only include
the simultaneous spikes generated by the MIP, but also include coincident
spikes within the uncorrelated presynaptic population.

\subsection{Hardware variations}

In contrast to arithmetic operations in software models, analog circuits
vary due to the manufacturing process, although they are identically
designed. The choice of precision for all building blocks should be
governed by those that distort network functionality most. In this
study, we assume that variations within the analog measurement and
accumulation circuits are likely to be a key requirement for these
choices, as they operate on the lowest level of STDP. Circuit variations
are measured and compared between the causal and anti-causal part
within a synapse and between synapses. All measurements are carried
out with the FACETS chip-based hardware system \citep{Schemmel06_1,Schemmel07_iscas}
with hardware parameters listed in \prettyref{tab:Appendix-Hardware-parameters}.
The FACETS chip-based hardware system shares a conceptually nearly
identical STDP circuit with the FACETS wafer-scale hardware system
(for details see \prettyref{sub:Appendix-STDP-chip}) which was still
in the assembly process at the course of this study. The hardware
measurements are written in PyNN \citep{Davison09} and use the workflow
described in \citet{Bruederle11_263}.

\subsubsection{Measurement\label{sub:Methods-Measurement}}

The circuit variations due to production imperfection are measured
by recording \textit{STDP curves} and comparing their integrals for
$\Delta t>0$ and $\Delta t<0$. The curves are recorded by applying
equidistant pairs of pre- and postsynaptic spikes with a predefined
latency $\Delta t$. Presynaptic spikes can be fed into the hardware
precisely. However, in contrast to NEST's parrot neurons, postsynaptic
spikes are not directly adjustable and therefore have to be evoked
by several synchronous external triggers (for details see \prettyref{sub:Appendix-Generating-spike-pairs}).
After discarding the first $10$ spike pairs to ensure regular firing,
the pre- and postsynaptic spike trains are shifted until the desired
latency $\Delta t$ is measured. Due to the low spike pair frequency
of $10\Hz$, only the correlations within and not between the spike
pairs are accumulated. The number $N$ of consecutive spike pairs
is increased until the threshold is crossed and hence a correlation
flag is set (\prettyref{fig:hw_variance}A). The inverse of this number
versus $\Delta t$ is called an STDP curve. Such curves were recorded
for $252$ synapses within one synapse column, the remaining $4$
synapses in this column were discarded.

For each STDP curve the total area $\aat=\aaa+\aac$ is calculated
and normalized by the mean $\overline{\aaabs}$ of the absolute area
$\aaabs=|\aaa|+|\aac|$ over all STDP curves. Ideally, $\aat$ would
vanish if both circuits are manufactured identically. The standard
deviation $\sigmaa$ (assuming Gaussian distributed measurement data)
of these normalized total areas $\aat$ is taken as one measure for
circuit variations. Besides this asymmetry which measures the variation
\textit{within} a synapse, a measure for variation \textit{across}
synapses is the standard deviation $\sigmat$ of the absolute areas
$\aaabs$. Therefore the absolute areas $\aaabs$ under each STDP
curve are again normalized by $\overline{\aaabs}$ and furthermore
the mean of all these normalized absolute areas is subtracted.

\subsubsection{Software analysis\label{sub:Methods-HW-Variation-Software}}

In order to predict the effects of the previously measured variations
on the network benchmark, these variations are integrated into computer
simulations. The thresholds for the causal and anti-causal spike pair
accumulations are drawn from two overlaying Gaussian distributions
defined by the ideal thresholds (\prettyref{eq:threshold}) and their
variations $\sigmat$, $\sigmaa$. Again, the same network benchmark
as described above is used, but with a fixed correlation coefficient
of $c=0.025$ and an 8-bit LUT configured with $n=12$ SSPs.

\clearpage{}

\section{Results}

Synaptic weights of the FACETS wafer-scale hardware system \citep{Schemmel10_1947}
have a 4-bit resolution. We show that such a weight resolution is
enough to exhibit learning in a neural network benchmark for synchrony
detection. To this end, we analyze the effects of weight discretization
in three steps as summarized in \prettyref{tab:Results-Outline}.

\begin{table}[H]
\mbox{\hspace{-1.3cm}\renewcommand{\arraystretch}{1.25}%
\begin{tabular}{>{\raggedright}p{10cm}|>{\raggedright}p{3.5cm}|>{\raggedright}p{3.5cm}}
\textbf{Description} & \textbf{Results} & \textbf{Methods}\tabularnewline
\hline 
\textbf{Look-up table analysis:}

Basic analyses on the configuration of STDP on discrete weights by
means of look-up tables (A) & A) \prettyref{sub:Results-Dynamic-Range} & A) \prettyref{sub:Methods-Discretized-Weights} and \ref{sub:Methods-Look-Up-Tables}\tabularnewline
and their long-term dynamics (B). & B) \prettyref{sub:Results-Equilibrium-Distributions} & B) \prettyref{sub:Methods-Equilibrium-Distributions}\tabularnewline
\hline 
\textbf{Spiking network benchmarks:}

Software implementation of hardware-inspired synapses with discrete
weights for application in spiking neural environments (C). &  & C) \prettyref{sub:Methods-Implementation-NEST}\tabularnewline
Analyses of their effects on short-term weight dynamics in single
synapses (D) & D) \prettyref{sub:Results-Whole-Synapse-Dynamics} & D) \prettyref{sub:Methods-Single-Synapse}\tabularnewline
and neural networks (E). & E) \prettyref{sub:Results-Synchrony-Detection} & E) \prettyref{sub:Methods-Network-Example}\tabularnewline
Analyses on how additional hardware constraints effect the network
benchmark (F). & F) \prettyref{sub:Results-Additional-Constraints} & F) \prettyref{sub:Methods-Network-Example}\tabularnewline
\hline 
\textbf{Hardware measurements:}

Measurement of hardware variations (G)  & G) \prettyref{sub:Results-Synapse-Variations} & G) \prettyref{sub:Methods-Measurement}\tabularnewline
and computer simulations analyzing their effects on the network benchmark
(H). & H) \prettyref{sub:Results-Synapse-Variations} & H) \prettyref{sub:Methods-HW-Variation-Software}\tabularnewline
\end{tabular}}

\caption{Outline of analyses on the effects of weight discretization and further
hardware constraints.\label{tab:Results-Outline}}
\end{table}

\subsection{Dynamic range of STDP on discrete weights\label{sub:Results-Dynamic-Range}}

We choose the configuration of STDP on discrete weights according
to \prettyref{sub:Methods-Discretized-Weights} and \prettyref{sub:Methods-Look-Up-Tables}
to obtain weight dynamics comparable to that in continuous weight
space. Each configuration can be described by a LUT ``projecting''
each discrete weight to new values, one for potentiation and one for
depression. For a given weight resolution $r$ the free configuration
parameter $n$ (number of SSPs) has to be adjusted to avoid a further
reduction of the usable weight resolution by \textit{dead discrete
weights}. Dead discrete weights are defined as weights projecting
to themselves in case of both potentiation and depression or not receiving
any projections from other discrete weights. The percentage of dead
discrete weights $d$ defines the lower and upper limit of feasible
values for $n$, the \textit{dynamic range}. The absolute value of
the interval within a SSP ($\deltats$) is an arbitrary choice merely
defining the granularity, but does not affect the results (not shown).
Note that spike timing precision in vivo, which is observed for high
dimensional input such as dense noise and natural scenes, goes rarely
beyond 5 to 10 ms \citep{Butts07_92,Desbordes08_324,Marre09_14596,Desbordes10_151,fregnac_cc2012},
and the choice of $10\ms$ as a granular step is thus justified biologically.

\begin{figure}
\mbox{\hspace{-1.3cm}

\includegraphics[width=18cm]{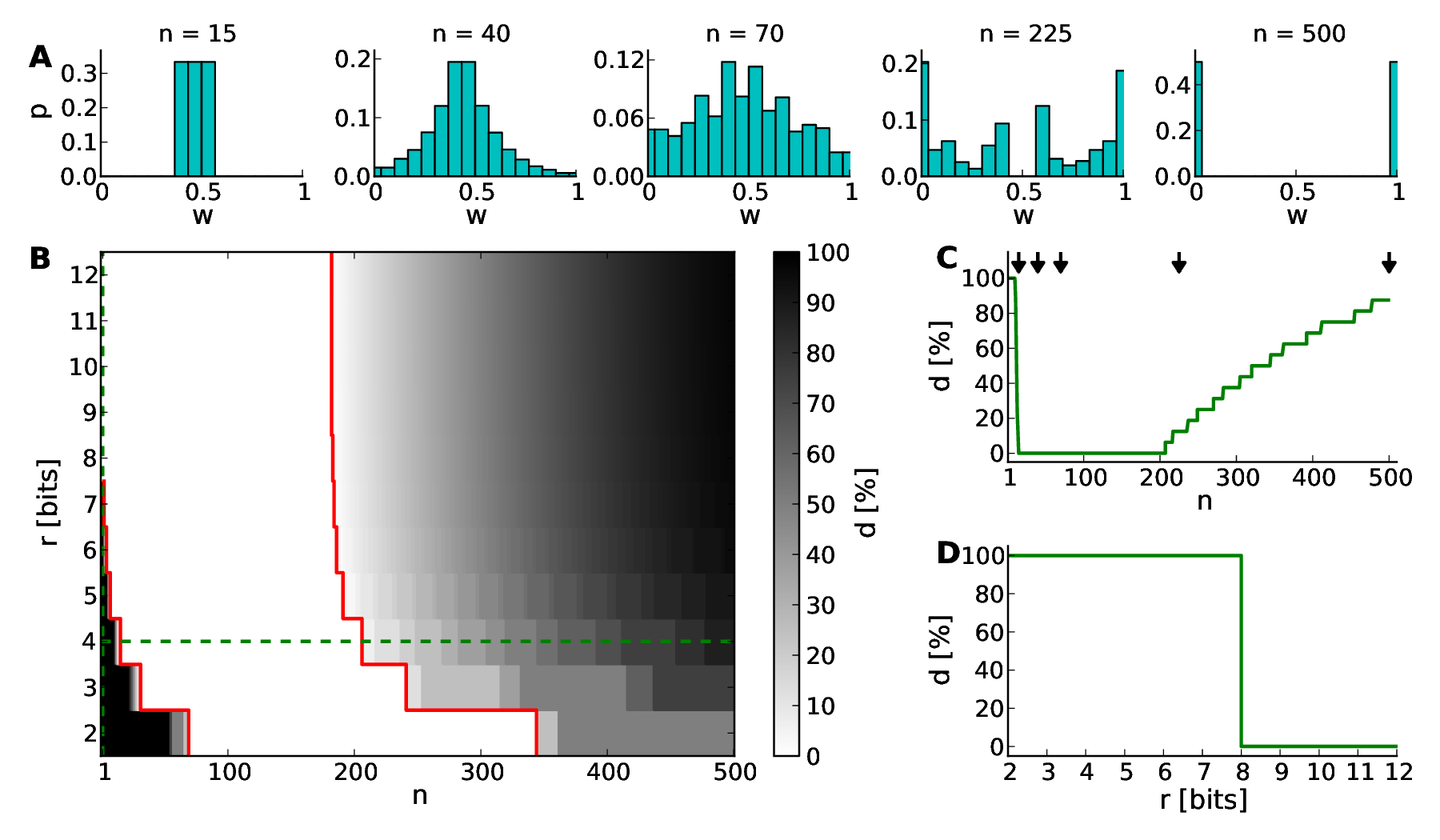}

}

\caption{The dynamic range for configurations of STDP on discrete weights.
\textbf{(A)} Equilibrium weight distributions for a 4-bit weight resolution:
Intermediate discrete weights partly project to themselves ($n=15$).
The equilibrium weight distribution widens with an increasing number
of SSPs ($n=40$ and $n=70$). For a large number of SSPs ($n=225$
and $n=500$) the intermediate discrete weights do not receive projections
from others. \textbf{(B) }Percentage of dead discrete weights $d$.
The limits of the dynamic range ($d=0\%$) are highlighted in red.
The limit towards low numbers of SSPs ($n=15$ in case of $r=4\bits$)
is caused by rounding effects (\prettyref{eq:conti-dis}), whereas
the upper limit ($n=206$ in case of $r=4\bits$) is caused by too
large weight updates. Green dashed lines indicate cross sections shown
in (C) and (D).\textbf{ (C)} Cross section of (B) at a 4-bit weight
resolution. The histograms shown in (A) are depicted with arrows.
\textbf{(D)} Cross section of (B) at $n=1$.\label{fig:LT_dynamic_range}}
\end{figure}

Generally, low values of $n$ realize frequent, small weight updates.
However, if $n$ is too low, some discrete weights may project to
themselves (see rounding in \prettyref{eq:conti-dis}) and prevent
synaptic weights from evolving dynamically (see \prettyref{tab:Example-LUTs-3}b
and $n=15$ in \prettyref{fig:LT_dynamic_range}A).

On the other hand, if $n$ exceeds the upper limit of the dynamic
range, intermediate discrete weights may not be reached by others.
Rare, large weight updates favor projections to discrete weights near
the borders of the weight range $I$ and lead to a bimodal equilibrium
weight distribution as shown in \prettyref{tab:Example-LUTs-3}c and
\prettyref{fig:LT_dynamic_range}A ($n=500$).

The lower limit of the dynamic range decreases with increasing resolution
(\prettyref{fig:LT_dynamic_range}B). Compared to a 4-bit weight resolution,
an 8-bit weight resolution is sufficiently high to resolve weight
updates down to a single SSP (\prettyref{fig:LT_dynamic_range}D).
This allows frequent weight updates comparable to weight evolutions
in continuous weight space. The upper limit of the dynamic range does
not change over increasing weight resolutions, but is critical for
limited update controller frequencies as investigated in \prettyref{sub:Results-Temporal-Evolution}.

\begin{table}
\begin{centering}
\renewcommand{\arraystretch}{1.25}\subfloat[]{\begin{centering}
\begin{tabular}{c|c|c}
$\wwd$ & $w_{+}$ & $w_{-}$\tabularnewline
\hline 
0 & $\frac{1}{3}$ & 0\tabularnewline
$\frac{1}{3}$ & $\frac{2}{3}$ & 0\tabularnewline
$\frac{2}{3}$ & 1 & $\frac{1}{3}$\tabularnewline
1 & 1 & $\frac{2}{3}$\tabularnewline
\end{tabular}
\par\end{centering}

}\quad{}\subfloat[]{\begin{centering}
\begin{tabular}{c|c|c}
$\wwd$ & $w_{+}$ & $w_{-}$\tabularnewline
\hline 
0 & $\frac{1}{3}$ & 0\tabularnewline
$\frac{1}{3}$ & $\frac{1}{3}$ & $\frac{1}{3}$\tabularnewline
$\frac{2}{3}$ & $\frac{2}{3}$ & $\frac{2}{3}$\tabularnewline
1 & 1 & $\frac{2}{3}$\tabularnewline
\end{tabular}
\par\end{centering}

}\quad{}\subfloat[]{\begin{centering}
\begin{tabular}{c|c|c}
$\wwd$ & $w_{+}$ & $w_{-}$\tabularnewline
\hline 
0 & $\frac{2}{3}$ & 0\tabularnewline
$\frac{1}{3}$ & 1 & 0\tabularnewline
$\frac{2}{3}$ & 1 & 0\tabularnewline
1 & 1 & 0\tabularnewline
\end{tabular}
\par\end{centering}

}
\par\end{centering}

\caption{Look-up tables for different numbers $n$ of SSPs. (a) As in \prettyref{tab:Example-LUTs}
($n=100$), which results in a LUT as expected. Weights are either
potentiated or depressed through the entire table. (b) $n=60$, which
is too low, because the discrete weights $\frac{1}{3}$ and $\frac{2}{3}$
are projecting exclusively to themselves. (c) $n=350$, which is too
large, because for $w_{+}$ the discrete weight $0$ is mapped right
to $\frac{2}{3}$ (and for $w_{-}$ the weight $1$ is mapped to $0$),
thus $\frac{1}{3}$ is never reached.\label{tab:Example-LUTs-3}}
\end{table}

\subsection{Equilibrium weight distributions\label{sub:Results-Equilibrium-Distributions}}

\begin{figure}
\mbox{\hspace{-1.3cm}

\includegraphics[width=18cm]{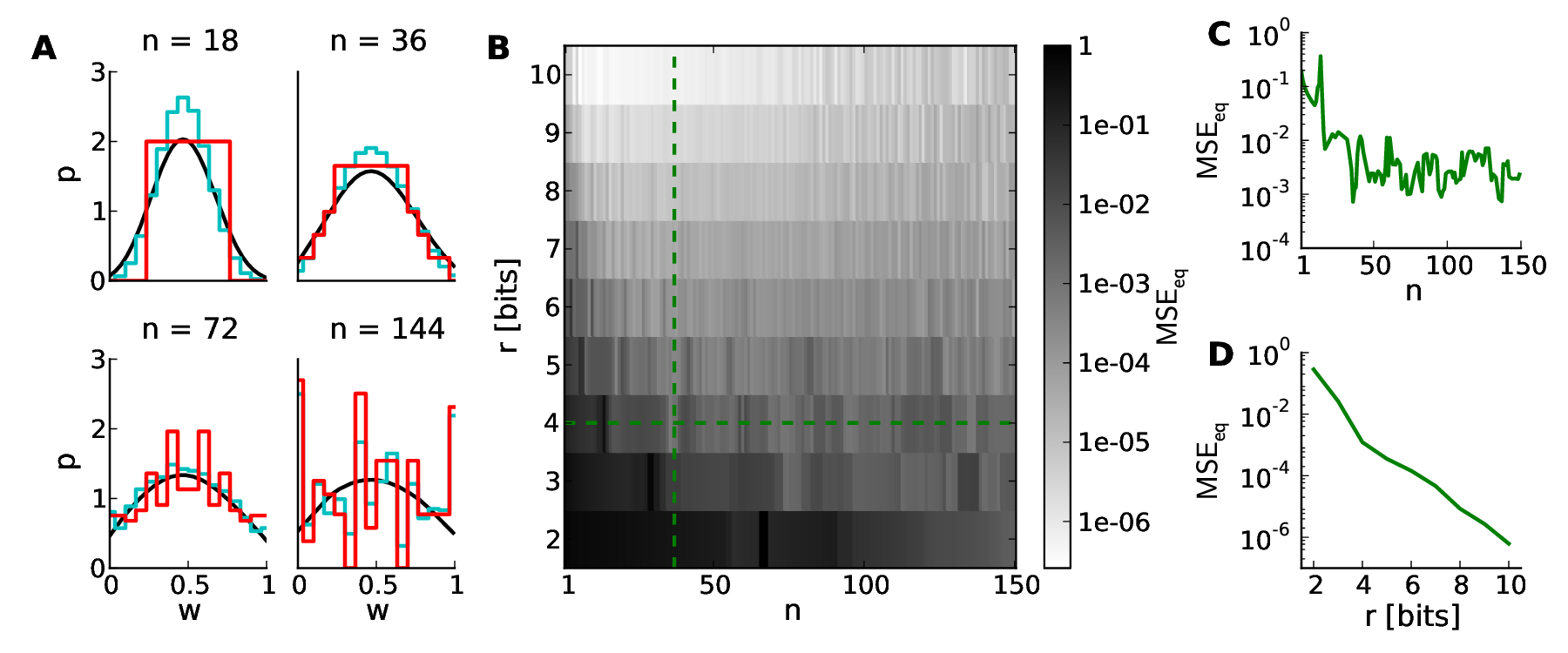}

}

\caption{Equilibrium weight distributions (long-term weight evolutions) for
configurations of STDP on discrete weights.\textbf{ (A)} Equilibrium
weight distributions for weight resolutions of $r=4\bits$ (red) and
$r=16\bits$ (cyan). Both distributions are displayed in 4-bit sampling,
for better comparison. Black curves depict the analytical approach.
We have chosen\textbf{ }$j=10^{5}$ iterations for generating each
discrete weight distribution to ensure convergence to the equilibrium
state. \textbf{(B)} Mean squared error $MSE_{eq}$ between the equilibrium
weight distributions for weight resolutions $r$ and the reference
weight resolution of 16 bits versus the number $n$ of SSPs. \textbf{(C),(D)}
Cross sections of (B) at $r=4\bits$ and $n=36$, respectively.\label{fig:equilibrium_distribution}}
\end{figure}

Studying learning in neural networks may span long periods of time.
Therefore we analyze equilibrium weight distributions being the temporal
limit of Poissonian distributed pre- and postsynaptic spiking. These
distributions are obtained by applying random walks on LUTs with uniformly
distributed occurrences of potentiations and depressions (\prettyref{sub:Methods-Equilibrium-Distributions}).
\prettyref{fig:equilibrium_distribution}A shows i.a.\ boundary effects
caused by LUTs configured within the upper part of the dynamic range.
E.g.\ for $n=144$, the relative frequencies of both boundary values
are increased due to large weight steps (red and cyan distributions).
Frequent weights, in turn, increase the probability of weights to
which they project (according to the LUT). This effect decreases with
the number of look-ups, due to the random nature of the stimulus,
however, causing intermediate weight values to occur at higher probability.

The impact of weight discretization on long-term weight dynamics is
quantified by comparing equilibrium weight distributions between low
and high weight resolutions. Weight discretization involves distortions
caused by rounding effects for small $n$ (\prettyref{eq:conti-dis}
and \prettyref{fig:LT_dynamic_range}) and boundary effects for high
$n$ (\prettyref{fig:equilibrium_distribution}A and C). High weight
resolutions can compensate for rounding effects, but not for boundary
effects (\prettyref{fig:equilibrium_distribution}B).

This analysis on long-term weight dynamics (\prettyref{fig:equilibrium_distribution}C)
refines the choice for $n$ roughly estimated by the dynamic range
(\prettyref{fig:LT_dynamic_range}C).

\subsection{Spiking network benchmarks\label{sub:Results-Temporal-Evolution}}

\begin{figure}
\mbox{\hspace{-1.3cm}

\includegraphics[width=18cm]{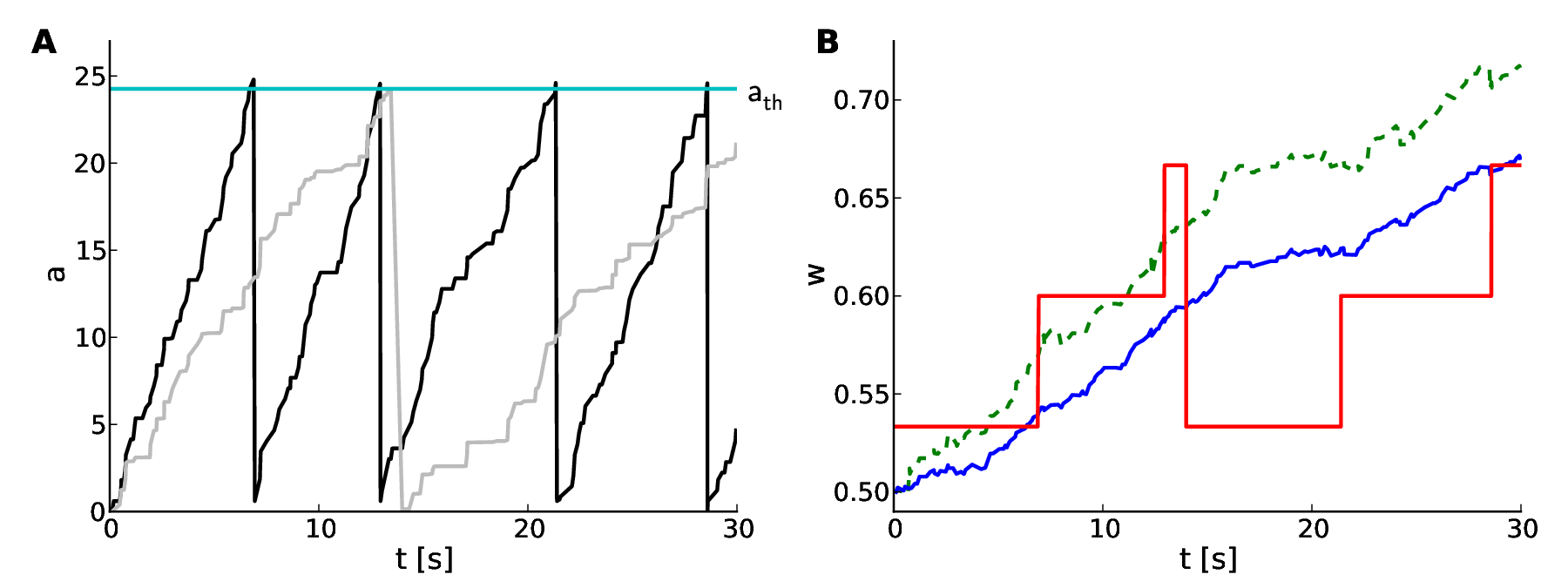}

}

\caption{Software implementation of STDP on discrete weights in spiking neural
networks.\textbf{ (A)} Temporal evolution of spike pair accumulations
$a$ (dimensionless) for causal (black) and anti-causal (gray) spike-timing-dependences.
If $a$ crosses the threshold $\ath$ (cyan), the weight is updated
and $a$ is reset to zero. Pre- and postsynaptic spike trains are
generated by a MIP with $c=0.5$ and $r=10\Hz$.\textbf{ (B)} Corresponding
weight evolution (solid red) for a 4-bit weight resolution and a LUT
configured with $n=30$. The weight evolution of the reference synapse
model with continuous weights, but a reduced symmetric nearest-neighbor
spike pairing scheme is depicted in solid blue. It differs from that
of a synapse model with continuous weights and an all-to-all spike
pairing scheme (dashed green).\label{fig:sums_and_traces}}
\end{figure}

We extend the above studies on temporal limits by analyses on short-term
dynamics with unequal probabilities for potentiation $\pp$ and depression
$\pd$. A hardware-inspired synapse model is used in computer simulations
of spiking neural networks, of which an example of typical dynamics
is shown in \prettyref{fig:sums_and_traces}. As the pre- and postsynaptic
spike trains are correlated in a causal fashion, the causal spike
pair accumulation increases faster than the anti-causal one (\prettyref{fig:sums_and_traces}A).
It crosses the threshold twice, evoking two potentiation steps (at
around $7\s$ and $13\s$) before the anti-causal spike pair accumulation
evokes a depression at around $14\s$ (\prettyref{fig:sums_and_traces}A
and B). The first two potentiations project to the subsequent entry
of the LUT, whereas the following depression rounds to the next but
one discrete weight (omitting one entry in the LUT) due to the asymmetry
measure $\alpha$ in the STDP model by \citet{Guetig03_3697}.

\subsubsection{Single synapse benchmark\label{sub:Results-Whole-Synapse-Dynamics}}

\begin{figure}
\mbox{\hspace{-1.3cm}

\includegraphics[width=18cm]{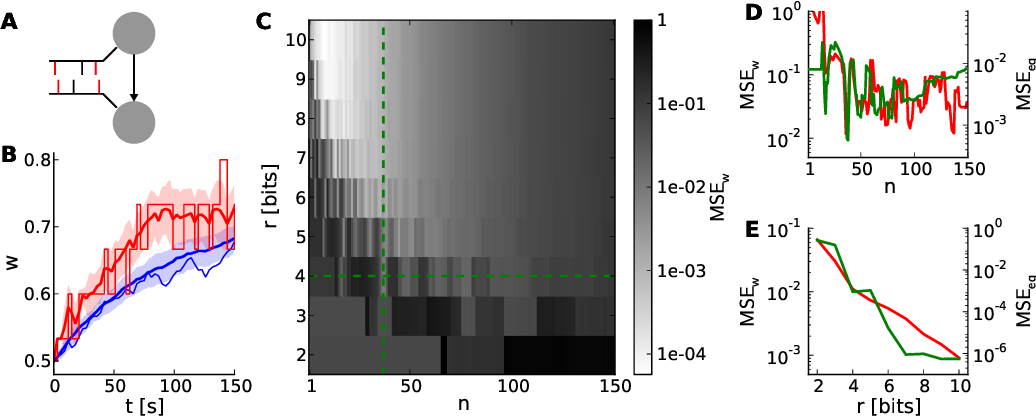}

}

\caption{Weight evolution of a single synapse with discrete weights. \textbf{(A)}
Network layout for single synapse analyses. An STDP synapse (arrow)
connects two neurons receiving correlated spike trains with correlation
coefficient $c$ (correlated spikes in red bars).\textbf{ (B)} Example
weight traces for the hardware-inspired ($r=4\bits$, $n=36$ in red)
and reference synapse model (blue). Means and standard deviations
over 30 realizations are plotted as bold lines and shaded areas, respectively.
The single weight traces for one arbitrarily chosen random seed are
depicted as thin lines. We applied a correlation coefficient $c=0.2$,
an initial weight $w_{0}=0.5$ and firing rates of $10\Hz$. The results
persist qualitatively for differing values staying within biologically
relevant ranges (not shown here). \textbf{(C)} Mean squared error
$MSE_{w}$ between the mean weight traces as shown in (A) over the
weight resolution $r$ and the number $n$ of SSPs. The parameters
$c$, $w_{0}$ and the firing rates are chosen as in (B). Other values
for $c$ and $w_{0}$ do not change the results qualitatively.\textbf{
(D),(E)} Cross sections of (C) at $r=4\bits$ and $n=36$ in green.
Red curves are adapted from \prettyref{fig:equilibrium_distribution}C
and D.\label{fig:single_synapse}}
\end{figure}

This benchmark compares single weight traces between hardware-inspired
and reference synapses (\prettyref{sub:Methods-Single-Synapse}).
A synapse receives correlated pre- and postsynaptic input (\prettyref{fig:single_synapse}A)
resulting in weight dynamics as shown in \prettyref{fig:single_synapse}B.
The standard deviation for discrete weights (hardware-inspired synapse
model) is larger than that for continuous weights (reference model).
This difference is caused by rare, large weight jumps (induced by
high $n)$ also responsible for the broadening of equilibrium weight
distributions (\prettyref{fig:equilibrium_distribution}A). Consequently,
the standard deviation increases further with decreasing weight resolutions
(not shown here).

The dependence of the deviation between discrete and continuous weight
traces on the weight resolution $r$ and the number $n$ of SSPs is
qualitatively comparable to that of comparisons between equilibrium
weight distributions (\prettyref{fig:single_synapse}D and E). This
similarity, especially in dependence on $n$ (\prettyref{fig:single_synapse}D),
emphasizes the crucial impact of LUT configurations on both short-
and long-term weight dynamics.

To further illustrate underlying rounding effects when configuring
LUTs, the asymmetry value $\alpha$ in Gütig's STDP model can be taken
as an example. In an extreme case both potentiation and depression
are rounded down (compare weight step size for potentiation and depression
in \prettyref{fig:sums_and_traces}B). This would increase the originally
slight asymmetry drastically and therefore enlarge the distortion
caused by weight discretization.

The weight update frequency $\nuw$ is determined by the weight resolution
$r$ and the number $n$ of SSPs. High frequencies are beneficial
for chronologically keeping up with weight evolutions in continuous
weight space. They can be realized by small numbers of SSPs lowering
the threshold $\ath$ (\prettyref{eq:threshold}). On the other hand,
rounding effects in the LUT configuration deteriorate for too small
numbers of SSPs (\prettyref{fig:single_synapse}D). In case of a weight
resolution $r=4\bits$ ($r=8\bits$) choosing $n=36$ ($n=12$) for
the LUT configuration represents a good balance between a high weight
update frequency and proper both short- and long-term weight dynamics
(\prettyref{fig:LT_dynamic_range}B, \prettyref{fig:equilibrium_distribution}B
and \prettyref{fig:single_synapse}C). Note that $n$ can be chosen
smaller for higher weight resolutions, because the distorting impact
of rounding effects decreases.

\subsubsection{Network benchmark: synchrony detection\label{sub:Results-Synchrony-Detection}}

Not only exact weight traces of single synapses (\prettyref{sub:Results-Whole-Synapse-Dynamics}),
but rather those of synapse populations are crucial to fulfill tasks,
e.g.\ the detection of synchronous firing within neural networks.
The principle of synchrony detection is a crucial feature of various
neural networks with plasticity, e.g.\ reported by \citet{Senn98_815,Kuba02_984,Davison06_5604,Boustani12_194}.
Here, it is introduced by means of an elementary benchmark neural
network (\prettyref{fig:synchrony_detector}A and \prettyref{sub:Methods-Network-Example}),
using the hardware-inspired or reference synapse model, respectively.

\prettyref{fig:synchrony_detector}B shows a delay distribution of
postsynaptic spike occurrences, relative to the trigger onset, synchronous
presynaptic firing (\prettyref{sub:Methods-Peri-Stimulus-Time-Histograms}).
For the shown range of $\deltatdel$, the postsynaptic neuron is more
likely to fire if connected with static (dark gray trace) instead
of STDP (black trace) synapses. The correlated population causes its
afferent synapses to strengthen more compared to those from the uncorrelated
population. This can be seen in \prettyref{fig:synchrony_detector}C,
where $w$ saturates at different values ($t\approx700\s$). The same
effect can be observed for discretized weights in \prettyref{fig:synchrony_detector}D.
For $\deltatdel>170\ms$ the delay distribution for static synapses
is larger than that for STDP synapses (not shown here), because such
delayed postsynaptic spikes are barely influenced by their presynaptic
counterparts. This is due to small time constants of the postsynaptic
neuron (see $\taum=\frac{C_{m}}{g_{L}}$ and $\tau_{syn}$ in \prettyref{tab:Appendix-Model-description}
and \ref{tab:Appendix-Simulation-parameters}) compared to $\deltatdel$.

\prettyref{fig:synchrony_detector}E shows the $p$-values of the
Mann-Whitney U test applied to both groups of synaptic weights at
$t=2,000\s$ for different configurations of weight resolution $r$
and number $n$ of SSPs. Generally, $p$-values (probability of having
the same median within both groups of weights) decrease with an increasing
correlation coefficient. Although applying previously selected ``healthy''
LUT configurations, weight discretization changes the required correlation
coefficient for reaching significance level (gray shaded areas). Incrementing
the weight resolution while retaining the number of SSPs $n$ does
not change the $p$-values significantly. Low weight resolutions cause
larger spacings between discrete weights that can further facilitate
the distinction between both medians (for $n=36$ compare $r=4\bits$
to $r=8\bits$ bits in \prettyref{fig:synchrony_detector}E). However,
reducing $n$ for high weight resolutions shortens the accumulation
period and consequently allows the synapses to capture fluctuations
in $a$ on smaller time scales. This improves the $p$-value, but
is inconvenient for low weight resolutions, because these LUT configurations
do not yield the desired weight dynamics (\prettyref{fig:LT_dynamic_range},
\ref{fig:equilibrium_distribution} and \ref{fig:single_synapse}).

\subsubsection{Network benchmark: further constraints\label{sub:Results-Additional-Constraints}}

In addition to the discretization of synaptic weights that has been
analyzed so far, we also consider additional hardware constraints
of the FACETS wafer-scale system (\prettyref{sub:Methods-Further-synapse-constraints}).
This allows us to compare the effects of other hardware constraints
to those of weight discretization.

\begin{figure}
\mbox{\hspace{-1.3cm}

\includegraphics[width=18cm]{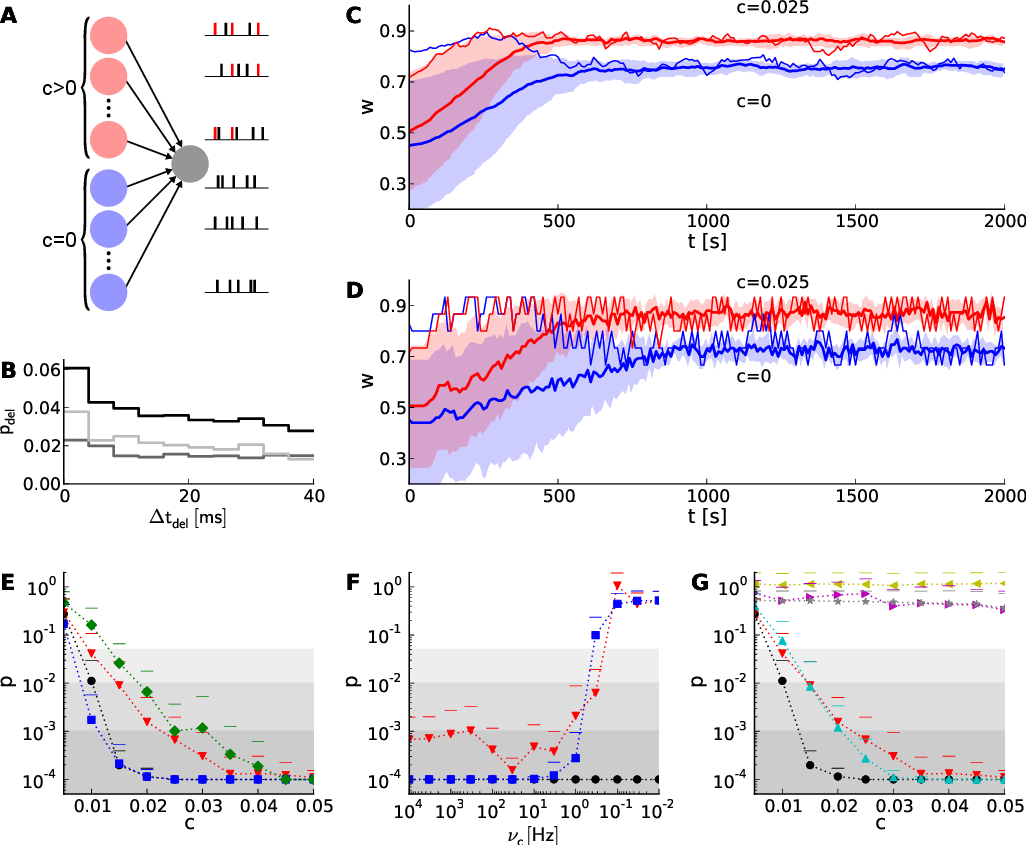}

}

\caption{Learning with discrete weights in a neural network benchmark for synchrony
detection. \textbf{(A)} Layout of the network benchmark. Two populations
of presynaptic neurons are connected to a postsynaptic neuron. On
the right, example spike trains of the presynaptic neurons are shown.
Red spikes indicate correlated firing due to shared spikes. \textbf{(B)}
PSTH for static synapses and STDP reference synapses. The light gray
histogram shows the difference between a simulation with STDP reference
synapses (black) and static synapses (dark gray). \textbf{(C)} The
mean weight traces (thick lines) and their standard deviations (shaded
areas) for both populations of afferent synapses using the reference
synapses model. Thin lines represent single synapses randomly chosen
for each population. \textbf{(D)} As in (B), but with the hardware-inspired
synapse model ($r=4\bits$ and $n=36$).\textbf{ (E)} The probability
($p$-value of Mann-Whitney U test) of having the same median of weights
within both groups of synapses (with correlated and uncorrelated input)
at $t=2,000\s$ versus the correlation coefficient $c$. The hardware-inspired
synapse model is represented in red ($r=4\bits$ and $n=36$), green
($r=8\bits$ and $n=36$) and blue ($r=8\bits$ and $n=12$). Black
depicts the reference synapse model ($r=64\bits$). The background
shading represents the significance levels: $p<0.05$, $p<0.01$ and
$p<0.001$.\textbf{ (F) }Dependence of the $p$-value on the update
controller frequency $\nuc$ for $c=0.025$. Colors as in (E) \textbf{(G)}
Black and red trace as in (E). Additionally, $p$-values for hardware-inspired
synapses with common resets are plotted in yellow ($r=4\bits$ and
$n=36$) and magenta ($r=8\bits$ and $n=12$). Compensations with
ADCs are depicted in gray ($r=4\bits$ and $n=15$ to $45$ in steps
of $2$) and cyan ($r=8\bits$ and $n=1$ to $46$ in steps of $3$).\label{fig:synchrony_detector}}
\end{figure}

First, we take into account a limited update controller frequency
$\nuc$. \prettyref{fig:synchrony_detector}F shows that low frequencies
($<1\Hz$) distort the weight dynamics drastically and deteriorate
the distinction between correlated and uncorrelated inputs. Ideally,
a weight update would be performed whenever the spike pair accumulations
cross the threshold (\prettyref{fig:sums_and_traces}A). However,
these weight updates of frequency $\nuw$ are now limited to a time
grid with frequency $\nuc$. The larger the latency between a threshold
crossing and the arrival of the weight update controller, the more
likely this threshold is exceeded. Hence, the weight update is underestimated
and delayed. Low weight resolutions are less affected, because a high
ratio $\frac{\nuc}{\nuw}$ reduces threshold overruns and hence distortions.
This low resolution requires a high number of SSPs which in turn increases
the threshold $\ath$ (\prettyref{eq:threshold}) and thus the weight
update frequency $\nuw$.

Second, hardware-inspired synapses with the limitation to common reset
lines cease to discriminate between correlated and uncorrelated input
(\prettyref{fig:synchrony_detector}G, yellow and magenta traces).
A crossing of the threshold by one spike pair accumulation resets
the other (\prettyref{fig:sums_and_traces}) and suppresses its further
weight updates, leading to underestimation of synapses with less correlated
input.

To compensate for common resets we suggest ADCs that allow the comparison
of spike pair accumulations to multiple thresholds. Nevertheless,
ADCs compensate common resets only for high weight resolutions (\prettyref{fig:synchrony_detector}G).
Again, for low weight resolutions and hence high numbers of SSPs fluctuations
can not be taken into account (\prettyref{fig:synchrony_detector}G,
gray values). This is the case for a 4-bit weight resolution, whereas
a 8-bit weight resolution is high enough to resolve small fluctuations
down to single SSPs (\prettyref{fig:synchrony_detector}G, cyan values).
Each threshold has its own LUT configured with a number of SSPs that
matches the dynamic range (\prettyref{fig:LT_dynamic_range}). The
upper limit of $n$ is chosen according to the results of \prettyref{sub:Results-Equilibrium-Distributions}.
The update controller frequency is chosen to be low enough ($\nuc=0.2\Hz$)
to enable all thresholds to be hit.

\subsection{Hardware variations\label{sub:Results-Synapse-Variations}}

\begin{figure}
\mbox{\hspace{-1.3cm}

\includegraphics[width=18cm]{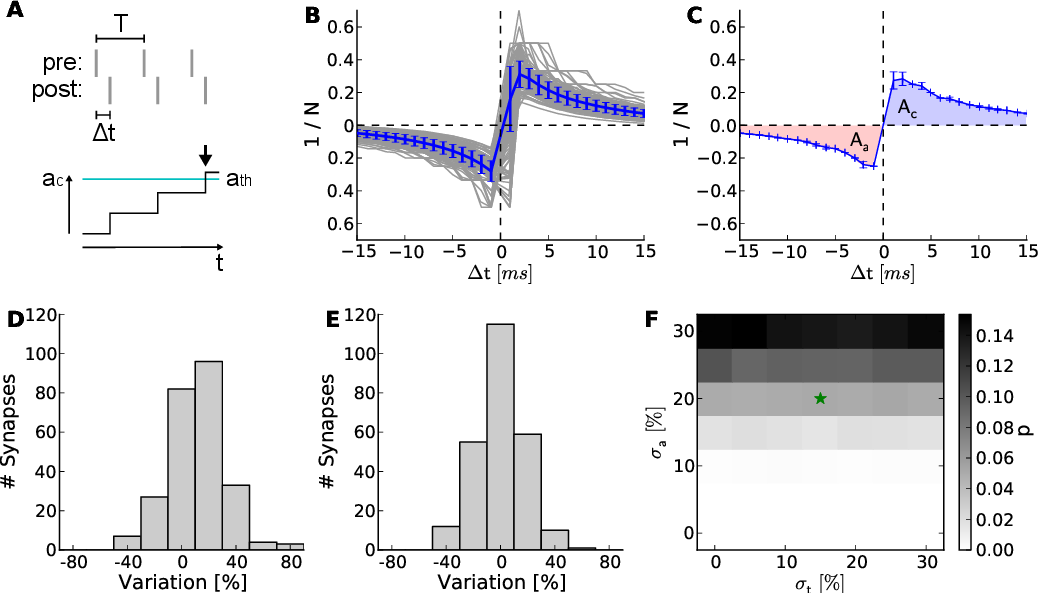}

}

\caption{Measurement of hardware synapse variations and their effects on learning
in the neural network benchmark. \textbf{(A)} Setup for recording
STDP curves. At the top, spike trains of the pre- and postsynaptic
neuron. Spike pairs with latency $\Delta t$ are repeated with frequency
$\frac{1}{T}$. At the bottom, a spike pair accumulation that crosses
the threshold $\ath$ (arrow). The inverse of the number of SSPs until
crossing $\ath$ (here $N=3$) is plotted in (B). \textbf{(B)} STDP
curves of $252$ hardware synapses within one synapse column (gray)
and their mean with error (blue).\textbf{ }A speed-up factor of $10^{5}$
is assumed. These curves correspond to $x(\Delta t)$ in \prettyref{eq:weight_time_separation},
whereas $F(w)$ is realized by the LUT. \textbf{(C)} One arbitrarily
chosen STDP curve (over $5$ trials) showing the areas for $\Delta t<0$
($\aaa$ in red) and $\Delta t>0$ ($\aac$ in blue). \textbf{(D)}
Asymmetry between $\aaa$ and $\aac$ within synapses ($\sigmaa=21\%$).
\textbf{(E)} Variation of the absolute areas between synapses ($\sigmat=17\%$).
\textbf{(F)} The $p$-value (as in \prettyref{fig:synchrony_detector}E-G)
in dependence on $\sigmaa$ and $\sigmat$. The values for (D) and
(E) are marked with an asterisk.\label{fig:hw_variance}}
\end{figure}

So far, we neglected production imperfections in real hardware systems.
However, fixed pattern noise induced by these imperfections are a
crucial limitation on the transistor level and may distort the functionality
of the analog synapse circuit making higher weight resolutions unnecessary.
The smaller and denser the transistors, the larger the discrepancies
from their theoretical properties \citep{Pelgrom89_1433}. Using the
protocol illustrated in \prettyref{fig:hw_variance}A we recorded
STDP curves on the FACETS chip-based hardware system (\prettyref{fig:hw_variance}B,
C and \prettyref{sub:Methods-Measurement}). Variations within ($\sigmaa$)
and between ($\sigmat$) individual synapses are shown as distributions
in \prettyref{fig:hw_variance}D and E, both suggesting variations
at around 20\%. Both variations are incorporated into computer simulations
of the network benchmark (\prettyref{fig:synchrony_detector}A and
\prettyref{sub:Methods-HW-Variation-Software}) to analyze their effects
on synchrony detection. The $p$-value (as in \prettyref{fig:synchrony_detector}E-G)
rises with increasing asymmetry within synapses, but is hardly affected
by variations between synapses (\prettyref{fig:hw_variance}F). 

\clearpage{}

\section{Discussion}

\subsection{Configuration of STDP on discrete weights}

In this study, we demonstrate generic strategies to configure STDP
on discrete weights as e.g.\ implemented in neuromorphic hardware
systems. Resulting weight dynamics is critically dependent on the
frequency of weight updates that has to be adjusted to the available
weight resolution. Choosing a frequency within the dynamic range (\prettyref{fig:LT_dynamic_range})
is a prerequisite for the exploitation of discrete weight space ensuring
proper weight dynamics. Analyses on long-term dynamics using Poisson-driven
equilibrium weight distributions help to refine this choice (\prettyref{fig:equilibrium_distribution}).
The obtained configuration space is similar to that of short-term
dynamics, being the evolution of single synaptic weights (\prettyref{fig:single_synapse}).
This similarity confirms the crucial impact of the LUT configuration
on weight dynamics which is caused by rounding effects. Based on these
results, we have chosen two example LUT configurations ($r=4\bits$;
$n=36$ and $r=8\bits$; $n=12$) for further analysis, both realizable
on the FACETS wafer-scale hardware system. High weight resolutions
allow for higher frequencies of weight updates approximating the ideal
model, occasionally requiring several spike pairs to evoke a weight
update. Correspondingly, in associative pairing literature, a minimal
number of associations is required to detect functional changes (expressed
by the spiking or postsynaptic potential response) and varies from
studies to studies from a few to several tens \citep{Cassenaer2007_709,Cassenaer12_nature}.

Discretization not only affects the accuracy of weights, but also
broadens their equilibrium weight distributions (\prettyref{fig:equilibrium_distribution}),
which are actually shown to be narrow in large-scale neural networks
\citep{Morrison07_1437}. Furthermore, this broadening can distort
the functionality of neural networks, e.g.\ it deteriorates the distinction
between the two groups of weights (of synapses originating from the
correlated or uncorrelated population) within the network benchmark
(compare \prettyref{fig:synchrony_detector}C to D). On the other
hand, weight discretization can also be advantageous for synchrony
detection, if e.g.\ groups of weights separate due to large step
sizes between neighboring discrete weights (compare red and green
in \prettyref{fig:synchrony_detector}E).

In summary, these analyses of STDP on discrete weights are necessary
for obtaining appropriate configurations for a variety of STDP models
and weight resolutions.

\subsection{4-bit weight resolution}

Simulations of the network benchmark show that a 4-bit weight resolution
is sufficient to detect synchronous presynaptic firing significantly
(\prettyref{fig:synchrony_detector}). Groups of synapses receiving
correlated input strengthen and in turn increase the probability of
synchronous presynaptic activity to elicit postsynaptic spikes as
compared to static synapses (\prettyref{fig:synchrony_detector}B).
Thus, the weight distribution within the network reflects synchrony
within sub-populations of presynaptic neurons. Increasing the weight
resolution causes both weight distributions, for the correlated and
uncorrelated input, to narrow and separate from each other. Consequently,
an 8-bit resolution is sufficient to reproduce the $p$-values of
continuous weights with floating point precision (corresponds to discrete
weights with $r=64\bits$, \prettyref{fig:synchrony_detector}E).
This resolution requires the combination of two hardware synapses
and is under development \citep{Schemmel10_1947}. On the other hand,
increasing the weight resolution, but retaining the frequency of weight
updates (number of SSPs), results in weight distributions of comparable
width and consequently does not improve the $p$-values significantly
(\prettyref{fig:synchrony_detector}E).

Other neuromorphic hardware systems implement bistable synapses corresponding
to a 1-bit weight resolution \citep{Badoni06_4,Indiveri10_1951}.
Bistable synapse models are shown to be sufficient for memory formation
\citep{Amit94_957,Fusi05_599,Brader07,Clopath08}. However, these
models do not only employ spike timings \citep{Levy83,Markram97a,Bi01,Mu2006_115,Cassenaer2007_709},
but also read the postsynaptic membrane potential \citep{Sjostrom01,Trachtenberg02_788}
requiring additional hardware resources. So far, there is no consensus
of a general synapse model, and neuromorphic hardware systems are
mostly limited to only subclasses of these models.

This studies on weight discretization are not limited to the FACETS
hardware systems only, but are applicable to other backends for neural
network simulations. For example, our results can be applied to the
fully digital neuromorphic hardware system described by \citet{Jin2010_1},
who also report STDP with a reduced weight resolution. Furthermore,
weight discretization may be a further approach to reduce memory consumption
of ``classical'' neural simulators.

\subsection{Further hardware constraints}

In addition to a limited weight resolution, we have studied further
constraints of the current FACETS wafer-scale hardware system with
the network benchmark.

A limited update controller frequency implying a minimum time interval
between subsequent weight updates does not affect the $p$-values
down to a critical frequency $\nuc\approx1\Hz$ (\prettyref{fig:synchrony_detector}F).
The update controller frequency decreases linearly with the number
of hardware synapses enabled for STDP. Assuming a hardware acceleration
factor of $10^{3}$ all synapses can be enabled for STDP staying below
this critical frequency. However, the number of STDP synapses should
be decreased if a higher update controller frequency is required,
e.g.\ for a configuration with an 8-bit weight resolution and a small
number of SSPs.

Common resets of spike pair accumulations reduce synapse chip resources
by requiring one instead of two reset lines, but suppress synaptic
depression and bias the weight evolution towards potentiation. This
is due to the feed-forward network architecture, in which causal relationships
between pre- and postsynaptic spikes are more likely than anti-causal
ones. Long periods of accumulation (large numbers of SSPs) lower the
probability of synaptic depression. Hence, all weights tend to saturate
at the maximum weight value impeding a distinction between both populations
of synapses within the network benchmark (\prettyref{fig:synchrony_detector}G).
The probability of synaptic depression can be increased by high weight
update frequencies (small numbers of SSPs) shortening the accumulation
periods (\prettyref{eq:threshold}) and subsequently approaching the
behavior of independent resets. However, high weight update frequencies
require high weight resolutions and thus high update controller frequencies,
which decreases the number of available synapses enabled for STDP.

As a compensation for common resets, we suggest that the single spike
pair accumulation threshold is expanded to multiple thresholds implemented
as ADCs. In comparison to synapses with common resets, ADCs improve
$p$-values significantly only for an 8-bit weight resolutions (\prettyref{fig:synchrony_detector}G,
compare cyan to magenta values). However, the combination of two 4-bit
hardware synapses allows to mimic independent resets and hence yields
$p$-values comparable to 8-bit synapses using ADCs (\prettyref{fig:synchrony_detector}G,
compare red to cyan values). Mimicking independent resets is under
development for the FACETS wafer-scale hardware system. Each of the
two combined synapses will be configured to accumulate only either
causal or anti-causal spike pairs, while both synapses are updated
in a common process. This requires only minor hardware design changes
within the weight update controller and should be preferred to more
expensive changes for realizing ADCs. The implementation of real second
reset lines is not possible without major hardware design changes,
but is considered for future chip revisions.

Benchmark simulations incorporating the measured variations within
and between synapse circuits due to production imperfections result
in $p$-values worse (higher) than for a 4-bit weight resolution (compare
asterisk in \prettyref{fig:hw_variance}F to red value for $c=0.025$
in \prettyref{fig:synchrony_detector}E). Consequently, a 4-bit weight
resolution is sufficient for the current implementation of the measurement
and accumulation circuits. We suppose that the isolatedly analyzed
effects of production imperfections and weight discretization add
up and limit the best possible $p$-value of each other. Analysis
on combinations of hardware restrictions would allow to quantify how
their effects add up and are considered for further studies. However,
hardware variations can also be considered as a limitation on the
transistor level making higher weight resolutions unnecessary.

\mbox{}

\begin{figure}
\begin{centering}
\includegraphics[width=85mm]{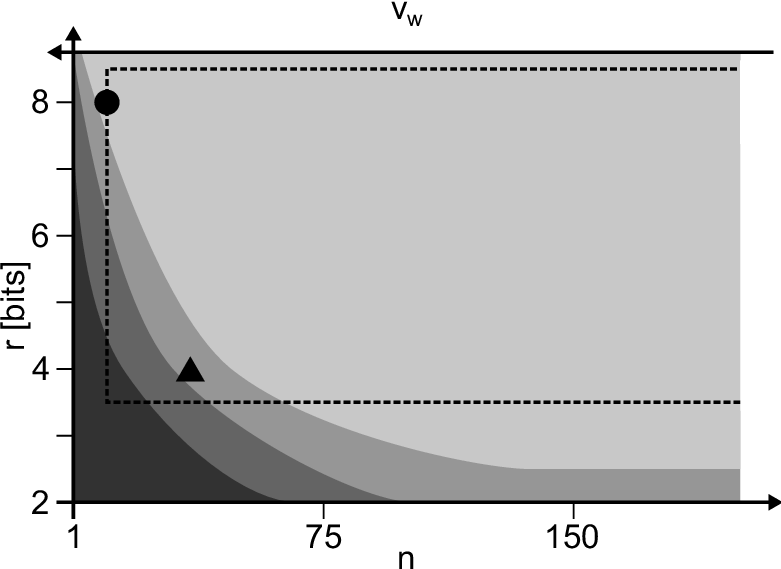}
\par\end{centering}

\caption{The configuration space of STDP on discrete weights spanned by the
weight resolution $r$ and the number $n$ of SSPs that is inversely
proportional to the weight update frequency $\nuw$. The darkest gray
area depicts the configurations with dead discrete weights (\prettyref{fig:LT_dynamic_range}).
The lower limits of configurations for proper equilibrium weight distributions
(\prettyref{fig:equilibrium_distribution}) and single synapse dynamics
(\prettyref{fig:single_synapse}) are shown with brighter shades.
The dashed rectangle marks configurations realizable by the FACETS
wafer-scale hardware system (assuming an acceleration factor of $10^{3}$,
all synapses enabled for STDP and SSPs applied with $10\Hz$). The
working points for a 4-bit ($n=36$) and 8-bit ($n=12$) weight resolution
are highlighted as a triangle and circle, respectively.\label{fig:The-nice-spots}}
\end{figure}

\prettyref{fig:The-nice-spots} summarizes the results on how to configure
STDP on discrete weights. For a given weight resolution $r$ the number
$n$ of SSPs has to be chosen as low as possible to allow for high
weight update frequencies $\nuw$. However, $n$ must be high enough
to ensure STDP dynamics comparable to continuous weights (lightest
gray shaded area) and to stay within the configuration space realizable
by the FACETS wafer-scale hardware system. The hardware system limits
the update controller frequency $\nuc$ and hence distorts STDP especially
for low $n$.

\subsection{Outlook}

Currently, STDP in neuromorphic hardware systems is enabled for only
$10$ to few $10,000$ synapses in real-time \citep{Arthur06_281,Zou06_211,Daouzli08_286,Ramakrishnan11_244}.
Large-scale systems do not implement long-term plasticity \citep{Morella06_4539,Vogelstein07_2281}
or operate in real-time only \citep{Jin10_91}. Enabling a large-scale
(over $4\cdot10^{7}$ synapses) and highly accelerated neuromorphic
hardware system (the FACETS wafer-scale hardware system) with configurable
STDP requires trade-offs between number and size of synapses, which
raises constraints in their implementation \citep{Schemmel06_1,Schemmel10_1947}.
\prettyref{tab:Hardware-synapse-reductions} summarizes these trade-offs
and gives an impression about the hardware costs and effects on STDP.

In this study, we introduced novel analysis tools allowing the investigation
of hardware constraints and therefore verifying and improving the
hardware design without the need for expensive and time-consuming
prototyping. Ideally, this validation process should be shifted to
an earlier stage of hardware design combining the expertise from Computational
Neuroscience and Neuromorphic Engineering, as e.g.\ published by
\citet{Linares-Barranco11}. This kind of research is crucial for
researchers to use and understand research executed on neuromorphic
hardware systems and thereby transform it into a tool substituting
von Neumann computers in Computational Neuroscience. \citet{Bruederle11_263}
report the development of a \textit{virtual hardware}, a simulation
tool replicating the functionality and configuration space of the
entire FACETS wafer-scale hardware system. This tool will allow further
analyses on hardware constraints, e.g.\ in the communication infrastructure
and configuration space.

The presented results verify the current implementation of the FACETS
wafer-scale hardware system in terms of balance between weight resolution,
update controller frequency and circuit variations. Further improvement
of the existing hardware implementation would require improvements
of all aspects. The only substantial bottleneck has been identified
to be common resets, already leading to design improvements of the
wafer-scale system.

Although all presented studies refer to the intermediate Gütig STDP
model, any other STDP model relying on \prettyref{eq:weight_time_separation}
and an exponentially decaying time-dependence can be investigated
with the existing software tools in a generic way, e.g.\ those models
listed in \prettyref{tab:stdp_models}. In contrast to the fixed exponential
time-dependence implemented as analog circuits in the FACETS wafer-scale
hardware system, the weight-dependence is freely programmable and
stored in a LUT.

Ideally, a high resolution in the weight range of highest plausibility
is requested, a high \textit{effective resolution}. Bounded STDP models
(e.g.\ the intermediate Gütig STDP model applied in this study) are
well suited for a 4-bit weight resolution and allow a linear mapping
of continuous to discrete weights. A 4-bit weight resolution causes
large weight updates and hence broadens the weight distribution spanning
the whole weight range. This results in a high effective resolution.
On the other hand, unbounded STDP models (e.g.\ the power law and
van Rossum STDP models) have long tails towards high weights. Cutting
the tail by only mapping low weights to discrete weights would increase
the frequency of the highest discrete weight. A possible solution
is a non-linear mapping of continuous to discrete weights - large
differences between high discrete weights and small differences between
low discrete weights. However, a variable distance between discrete
weights would require more hardware efforts.

An all-to-all spike pairing scheme applied to the reference synapses
within the network benchmark results in $p$-values worse (higher)
than for synapses implementing a reduced symmetric nearest-neighbor
spike pairing scheme (not shown, but comparable to 4-bit discrete
weights in \prettyref{fig:synchrony_detector}E, see red values).
Detailed analyses on different spike pairing schemes could be investigated
in further studies.

As a next step, our hardware synapse model can replace the regular
STDP synapses in simulations of established neural networks, to test
their robustness and applicability for physical emulation in the FACETS
wafer-scale hardware system. The synapse model is available in the
following NEST release and can easily be applied to NEST or PyNN network
descriptions. If neural networks, or modifications of them, qualitatively
reproduce the simulation, they can be applied to the hardware system,
with which similar results can be expected. Thus, the presented simulation
tools allow beforehand modifications of network architectures to ensure
the compatibility with the hardware system.

With respect to more complex long-term plasticity models, the hardware
system is currently being extended by a programmable microprocessor
that is in control of all weight modifications. This processor allows
to combine synapse rows in order to compensate for common resets.
With possible access to further neuron or network properties the processor
would allow for more complex plasticity rules as e.g.\ those of \citet{Clopath08}
and \citet{Vogels11_1569}. Even modifications of multiple neurons
are feasible, a phenomenon observed in experiments with neuromodulators
\citep{Eckhorn90,Itti2001_194,Reynolds02,Shmuel2005_2117}. Nevertheless,
more experimental data and consensus about neuromodulator models and
their applications are required to further customize the processor.
New hardware revisions are rather expensive and consequently should
only cover established models that are prepared for hardware implementation
by dedicated studies.

This presented evaluation of the FACETS wafer-scale hardware system
is meant to encourage neuroscientists to benefit from neuromorphic
hardware without leaving their environment in terms of neuron, synapse
and network models. We further endorse that, towards an efficient
exploitation of hardware resources, the design of synapse models will
be influenced by hardware implementations rather than only by their
mathematical treatability \citep[e.g.][]{Badoni06_4}.

\begin{table}
\begin{centering}
\begin{tabular}{>{\raggedright}p{5cm}|>{\centering}p{2cm}|>{\raggedright}p{4.5cm}}
Modification & Resource reduction & Effect on STDP\tabularnewline
\hline 
\hline 
Global\textbf{ }weight update controller & +++ & Latency between synapse processings; spike pair accumulations necessary\tabularnewline
\hline 
Analog\textbf{ }measurement of spike-timing-dependence & ++ & Analog measurements are affected by production imperfections\tabularnewline
\hline 
Reduced\textbf{ }spike pairing scheme & ++ & n.a.\tabularnewline
\hline 
Decreased\textbf{ }weight resolution & ++ & Loss in synapse dynamics and competition; large weight steps require
spike pair accumulations\tabularnewline
\hline 
Reduction of operation frequency\textbf{ }$\nuc$ of the weight update
controller (overall frequency could be increased by implementing multiple
controllers) & ++ & Threshold over-shootings distorts synchrony detection\tabularnewline
\hline 
Common reset\textbf{ }line & + & No synchrony detection possible\tabularnewline
\hline 
LUTs (compared to arithmetic operations) & + & None\tabularnewline
\hline 
ADCs as compensation for common resets & - & No significant compensation in case of 4-bit synapses\tabularnewline
\end{tabular}
\par\end{centering}

\caption{Possible design modifications of hardware synapses, their reduction
in terms of required chip resources and their effects on STDP. These
modifications are listed by their resource reduction in descending
order inspired by the FACETS wafer-scale hardware system and its production
process. A larger reduction of chip resources allows more synapses
on a single chip.\label{tab:Hardware-synapse-reductions}}
\end{table}

\section{Acknowledgment}

The research leading to these results has received funding by the
European Union 6th and 7th Framework Programme under grant agreement
no.\ 15879 (FACETS) and no.\ 269921 (BrainScaleS). Special thanks
to Yves Frégnac, Daniel Brüderle and Andreas Grübl for helpful discussions
and technical support.

\clearpage{}

\bibliographystyle{neuralcomput_natbib}
\bibliography{brain,math,computer,manual}
\clearpage{}

\section{Appendix}

\subsection{Analytical distributions\label{sub:Appendix-Analytical-Distributions}}

Weight evolutions can be described by asymmetric Markov processes
with boundary conditions. Following \citet{VanRossum00}, the weight
distribution $P(w)$ can be expressed by a Taylor expansion of the
underlying master equation

\begin{equation}
\frac{\partial P(w,t)}{\partial t}=-\pd P(w,t)-\pp P(w,t)+\pd P(w+\deltawd,t)+\pp P(w-\deltawp,t).\label{eq:master}
\end{equation}
In contrast to \citet{VanRossum00}, this study defines a weight step
$\Delta w$ by a sequence of $n$ weight updates $\delta w$ as described
by \prettyref{eq:weight_time_separation}. Hence the weight steps
$\Delta w$ can be written as $\deltawd(w)=(w+F_{-}(w))_{n}-w$ and
$\deltawp(w)=(w+F_{+}(w))_{n}-w$, where $f(w)_{n}$ is the $n$-th
recursive evaluation of $f(w)$. 

According to \citet{VanRossum00} this Taylor expansion results in
the Fokker-Planck equation
\begin{equation}
\frac{\partial P(w,t)}{\partial t}=-\frac{\partial}{\partial w}\left[A(w)P(w,t)\right]+\frac{1}{2}\frac{\partial^{2}}{\partial w^{2}}\left[B(w)P(w,t)\right]\label{eq:fokkerplanck}
\end{equation}
with jump moments $A(w)=\pd\deltawd(w)+\pp\deltawp(w)$ and $B(w)=\pd\deltawd(w)^{2}+\pp\deltawp(w)^{2}$,
which has the following solution for reflecting boundary conditions
\citep{Gardiner09}:
\begin{equation}
P(w)=\frac{N}{B(w)}\exp\left[2\int_{0}^{w}\frac{A(w')}{B(w')}dw'\right],\label{eq:analytical_integral}
\end{equation}
with $N$ as a normalization factor. For small $n$ this equation
can be solved analytically, but is integrated numerically to cover
also large $n$.

However, this analytical approach fails, because the Taylor expansion
in combination with the boundary conditions does not hold for large
$n$ (absorbing boundary conditions do not improve the results).

\subsection{STDP in the FACETS chip-based hardware system\label{sub:Appendix-STDP-chip}}

The STDP mechanism of the FACETS chip-based hardware system differs
from that of the FACETS wafer-scale hardware system as follows. The
major difference is the comparison of spike pair accumulations with
thresholds. The wafer-scale system analyzed in this study compares
both spike pair accumulations with a threshold (the threshold can
be set independently for both accumulations, but they are assumed
to be equal in this study). An weight update is performed if a single
accumulation crosses this threshold. In contrast, the chip-based system
used for all measurements subtracts both spike pair accumulations
and compares the absolute value of their difference $\left|\ac-\aa\right|$
with a single threshold. If this threshold is crossed, the sign of
the difference between the spike pair accumulations $\mathrm{sig}\left(\ac-\aa\right)$
determines, whether the causal or anti-causal accumulation prevails
and the weight is updated accordingly. However, this difference between
both hardware systems can be neglected, because both STDP mechanisms
are identical if exclusively causal or anti-causal spike pairs are
accumulated. This is the case for the measurement protocol of STDP
curves.

\subsection{Generating spike pairs in hardware\label{sub:Appendix-Generating-spike-pairs}}

Spike pairs in the FACETS chip-based hardware system are generated
as follows. Presynaptic spike times can be set precisely, whereas
postsynaptic spikes need to be triggered by presynaptic input. Therefore,
a presynaptic spike (via the measured synapse) and $m$ trigger spikes
(eliciting a postsynaptic spike) are fed into a single neuron occupying
$m+1$ synapses. The synaptic weights as well as the synapse driver
strengths of the trigger synapses are proportional to the synaptic
peak conductance and are adjusted in such a way that a single postsynaptic
spike is evoked. The highest reliability of spike times within a hardware
run and between runs is achieved for $m=4$ trigger synapses (not
shown here). The synapse driver strength is set to the intermediate
value between the limiting case of no and multiple postsynaptic spikes
evoked by one trigger only. The synaptic weight of the measured synapse
is set to zero and consequently the measured synapse has no influence
on the elicitation of postsynaptic spikes.

\begin{table}
\begin{centering}
\begin{tabular}{>{\centering}p{2.5cm}|>{\centering}p{6cm}|>{\centering}p{2cm}}
Parameter & Description & Value\tabularnewline
\hline 
\hline 
$V_{\mathrm{clrc}}$ & Amount of charge that will be accumulated on the capacitor $C_{1}$
(\citealp{Schemmel06_1}) in case of causal spike time correlations,
corresponds to $x(\Delta t)$ & $0.90\V$\tabularnewline
\hline 
$V_{\mathrm{clra}}$ & See $V_{\mathrm{clrc}}$, but for the anti-causal circuit & $0.94\V$\tabularnewline
\hline 
$V_{\mathrm{ctlow}}$ & Lower spike pair accumulation threshold & $0.85\V$\tabularnewline
\hline 
$V_{\mathrm{cthigh}}$ & Higher spike pair accumulation threshold & $1.0\V$\tabularnewline
\hline 
$adjdel$ & Adjustable delay between the pre- and postsynaptic spike & $2.5\uA$\tabularnewline
\hline 
$V_{\mathrm{m}}$ & Parameter to stretch the STDP time constant $\taustdp$ & $0.0\V$\tabularnewline
\hline 
$I_{\mathrm{bcorreadb}}$ & Bias current that influences timing issues during read outs & $2.0\uA$\tabularnewline
\hline 
$drvI_{\mathrm{rise}}$ & Rise time of synaptic conductance & $1.0\V$\tabularnewline
\hline 
$drvI_{\mathrm{fall}}$ & Fall time of synaptic conductance & $1.0\V$\tabularnewline
\hline 
$V_{\mathrm{start}}$ & Start value of synaptic conductance, need for small rise times & $0.25\V$\tabularnewline
\hline 
$drvI_{\mathrm{out}}$ & Maximum value of synaptic conductance, corresponds to $\gmax$ & variable\tabularnewline
\end{tabular}
\par\end{centering}

\caption{Applied hardware parameters. The difference $V_{\mathrm{cthigh}}-V_{\mathrm{ctlow}}$
corresponds to the threshold $\ath$. All data is recorded with the
FACETS chip-based hardware system using chip number $444$ and synapse
column $4$.\label{tab:Appendix-Hardware-parameters}}
\end{table}
\begin{table}
\begin{tabular}{|>{\raggedright}p{3cm}|>{\raggedright}p{12cm}|}
\hline 
\multicolumn{2}{|>{\color{white}\columncolor{black}}c|}{\textbf{A: Model summary}}\tabularnewline
\hline 
\textbf{Populations} & three: uncorrelated input (U), correlated input (C), target (T)\tabularnewline
\hline 
\textbf{Topology} & feed-forward\tabularnewline
\hline 
\textbf{Connectivity} & all-to-one\tabularnewline
\hline 
\textbf{Neuron model } & leaky integrate-and-fire, fixed voltage threshold, fixed absolute
refractory period (voltage clamp)\tabularnewline
\hline 
\textbf{Synapse model} & exponential-shaped postsynaptic conductances\tabularnewline
\hline 
\textbf{Plasticity} & intermediate Gütig spike-timing dependent plasticity\tabularnewline
\hline 
\textbf{Input} & fixed-rate Poisson (for U) and multiple interaction process (for C)
spike trains \tabularnewline
\hline 
\textbf{Measurements} & synaptic weights\tabularnewline
\hline 
\end{tabular}

\begin{tabular}{|>{\raggedright}p{3cm}|>{\raggedright}p{3cm}|>{\raggedright}p{8.57cm}|}
\hline 
\multicolumn{3}{|>{\color{white}\columncolor{black}}c|}{\textbf{B: Populations}}\tabularnewline
\hline 
\textbf{Name} & \textbf{Elements} & \textbf{Population size}\tabularnewline
\hline 
U & parrot neurons & $\nnu$\tabularnewline
\hline 
C & parrot neurons & $\nnc$\tabularnewline
\hline 
T & iaf neurons & $\nnt$\tabularnewline
\hline 
\end{tabular}

\begin{tabular}{|>{\raggedright}p{3cm}|>{\raggedright}p{3cm}|>{\raggedright}p{8.57cm}|}
\hline 
\multicolumn{3}{|>{\color{white}\columncolor{black}}c|}{\textbf{C: Connectivity}}\tabularnewline
\hline 
\textbf{Source} & \textbf{Target} & \textbf{Pattern}\tabularnewline
\hline 
U & T & \multirow{2}{8.57cm}{all-to-all, uniformly distributed initial weights $w$, STDP, delay
$d$}\tabularnewline
\cline{1-2} 
C & T & \tabularnewline
\hline 
\end{tabular}

\begin{tabular}{|>{\raggedright}p{3cm}|>{\raggedright}p{12cm}|}
\hline 
\multicolumn{2}{|>{\color{white}\columncolor{black}}c|}{\textbf{D: Neuron and synapse model}}\tabularnewline
\hline 
\textbf{Name} & iaf neuron\tabularnewline
\hline 
\textbf{Type} & leaky integrate-and-fire, exponential-shaped synaptic conductances\tabularnewline
\hline 
\textbf{Sub-threshold dynamics} & $\cm\frac{\mathrm{d}V}{\mathrm{d}t}=\gl(\el-V)+g(t)(\ee-V)$ if $t>t^{*}+\taur$

$V(t)=\vreset$ else

$g(t)=w\gmax\exp(-t/\taus)$\tabularnewline
\hline 
\textbf{Spiking} & If $V(t-)<\theta\wedge V(t+)\geq\theta$

1. set $t^{*}=t$, 2. emit spike with time stamp $t^{*}$\tabularnewline
\hline 
\textbf{Name} & parrot neuron\tabularnewline
\hline 
\textbf{Type} & repeats input spikes with delay $d$\tabularnewline
\hline 
\end{tabular}

\begin{tabular}{|>{\raggedright}p{3cm}|>{\raggedright}p{12cm}|}
\hline 
\multicolumn{2}{|>{\color{white}\columncolor{black}}c|}{\textbf{E: Plasticity}}\tabularnewline
\hline 
\textbf{Name} & intermediate Gütig STDP\tabularnewline
\hline 
\textbf{Spike pairing scheme} & reduced symmetric nearest-neighbor\tabularnewline
\hline 
\textbf{Weight dynamics} & $\delta w(w,\Delta t)=F(w)x(\Delta t)$

$x(\Delta t)=\exp(-|\Delta t|/\taustdp)$

$F(w)=\lambda(1-w)^{\mu}$ if $\Delta t>0$

$F(w)=-\lambda\alpha w^{\mu}$ if $\Delta t<0$\tabularnewline
\hline 
\end{tabular}

\begin{tabular}{|>{\raggedright}p{3cm}|>{\raggedright}p{3cm}|>{\raggedright}p{8.57cm}|}
\hline 
\multicolumn{3}{|>{\color{white}\columncolor{black}}c|}{\textbf{F: Input}}\tabularnewline
\hline 
\textbf{Type} & \textbf{Target} & \textbf{Description}\tabularnewline
\hline 
Poisson generators & U & independent Poisson spike trains with firing rate $\rho$\tabularnewline
\hline 
MIP generators & C & spike trains with correlation $c$ and firing rate $\rho$\tabularnewline
\hline 
\end{tabular}

\selectlanguage{american}%
\begin{tabular}{|>{\raggedright}p{15.43cm}|}
\hline 
\multicolumn{1}{|>{\color{white}\columncolor{black}}c|}{\selectlanguage{english}%
\textbf{G: Measurements}\selectlanguage{american}%
}\tabularnewline
\hline 
\selectlanguage{english}%
evolution and final distribution of all synaptic weights\selectlanguage{american}%
\tabularnewline
\hline 
\end{tabular}

\selectlanguage{english}%
\caption{Model description of the network benchmark using the reference synapse
model. After \citet{Nordlie-2009_e1000456}. For details about the
hardware-inspired synapse model see \prettyref{sub:Methods-Implementation-NEST}.\label{tab:Appendix-Model-description}}
\end{table}
\begin{table}
\begin{tabular}{|>{\raggedright}p{3cm}|>{\raggedright}p{3cm}|>{\raggedright}p{8.57cm}|}
\hline 
\multicolumn{3}{|>{\columncolor{parametergray}}c|}{\textbf{B: Populations}}\tabularnewline
\hline 
\textbf{Name} & \textbf{Value} & \textbf{Description}\tabularnewline
\hline 
$\nnu$ & $10$ & number of neurons in uncorrelated input population\tabularnewline
\hline 
$\nnc$ & $10$ & number of neurons in correlated input population\tabularnewline
\hline 
$\nnt$ & $1$ & number of neurons in target population\tabularnewline
\hline 
\end{tabular}

\begin{tabular}{|>{\raggedright}p{3cm}|>{\raggedright}p{3cm}|>{\raggedright}p{8.57cm}|}
\hline 
\multicolumn{3}{|>{\columncolor{parametergray}}c|}{\textbf{C: Connectivity}}\tabularnewline
\hline 
\textbf{Name} & \textbf{Value} & \textbf{Description}\tabularnewline
\hline 
$w$ & uniformly distributed over {[}0,1{]} & number of neurons in uncorrelated input population\tabularnewline
\hline 
$d$ & $0.1\ms$ & synaptic transmission delays\tabularnewline
\hline 
\end{tabular}

\begin{tabular}{|>{\raggedright}p{3cm}|>{\raggedright}p{3cm}|>{\raggedright}p{8.57cm}|}
\hline 
\multicolumn{3}{|>{\columncolor{parametergray}}c|}{\textbf{D: Neuron and synapse model}}\tabularnewline
\hline 
\textbf{Name} & \textbf{Value} & \textbf{Description}\tabularnewline
\hline 
$C_{\mathrm{m}}$ & $250\pF$ & membrane capacity\tabularnewline
\hline 
$\gl$ & $16.6667\nS$ & leakage conductance\tabularnewline
\hline 
$\el$ & $-70\mV$ & leakage reversal potential\tabularnewline
\hline 
$\theta$ & $-55\mV$ & fixed firing threshold\tabularnewline
\hline 
$\vreset$ & $-60\mV$ & reset potential\tabularnewline
\hline 
$\taur$ & $2\ms$ & absolute refractory period\tabularnewline
\hline 
$\ee$ & $0\mV$ & excitatory reversal potential\tabularnewline
\hline 
$\gmax$ & $100\nS$ & postsynaptic maximum conductance\tabularnewline
\hline 
$\taus$ & $0.2\ms$ & postsynaptic conductance time constant\tabularnewline
\hline 
\end{tabular}

\begin{tabular}{|>{\raggedright}p{3cm}|>{\raggedright}p{3cm}|>{\raggedright}p{8.57cm}|}
\hline 
\multicolumn{3}{|>{\columncolor{parametergray}}c|}{\textbf{E: Plasticity}}\tabularnewline
\hline 
\textbf{Name} & \textbf{Value} & \textbf{Description}\tabularnewline
\hline 
$\alpha$ & $1.05$ & asymmetry\tabularnewline
\hline 
$\lambda$ & $0.005$ & learning rate\tabularnewline
\hline 
$\mu$ & $0.4$ & exponent\tabularnewline
\hline 
$\taustdp$ & $20\ms$ & STDP time constant\tabularnewline
\hline 
\end{tabular}

\begin{tabular}{|>{\raggedright}p{3cm}|>{\raggedright}p{3cm}|>{\raggedright}p{8.57cm}|}
\hline 
\multicolumn{3}{|>{\columncolor{parametergray}}c|}{\textbf{F: Input}}\tabularnewline
\hline 
\textbf{Name} & \textbf{Value} & \textbf{Description}\tabularnewline
\hline 
$\rho$ & $7.2\Hz$ & firing rate\tabularnewline
\hline 
$c$ & {[}0.005,0.05{]} & pair-wise correlation between spike trains\tabularnewline
\hline 
\end{tabular}

\caption{Parameter specification. The categories refer to the model description
in \prettyref{tab:Appendix-Model-description}.\label{tab:Appendix-Simulation-parameters}}
\end{table}

\end{document}